\documentclass[aps,preprint,showpacs,superscriptaddress,groupedaddress]{revtex4-1}  
\usepackage{graphicx}  
\usepackage{dcolumn}   
\usepackage{bm}        
\usepackage{amssymb}   
\usepackage{amsmath}   
\usepackage{listings}
\usepackage{array}
\usepackage{graphicx}
\usepackage{lineno}
\usepackage{dcolumn}
\usepackage{color}
\usepackage{overpic}
\usepackage{multirow}
\usepackage{hyperref}

\newcommand{\mL}{\mathcal{L}}
\newcommand{\hatmu}{\hat{\mu}}

\newcommand{\hatthe}{\hat{\theta}}

\newcommand{\hhatthe}{\hat{\hat{\theta}}}

\newcommand{\vx}{\vec{x}}
\newcommand{\vtheta}{\vec{\theta}}
\newcommand{\etmiss}{E_{\text{T}}^{\text{miss}}}
\newcommand{\MET}{E_{\text{T}}^{\text{miss}}}
\newcommand{\tauhad}{\tau_{\text{had}}}
\newcommand{\Nbins}{N_{\text{bins}}}
\newcommand{\Nsysts}{N_{\text{systs}}}
\newcommand{\Nevents}{N_{\text{events}}}
\newcommand{\ID}{\text{ID}}
\newcommand{\pT}{p_{\text{T}}}

\begin{document}

\widetext


\title{QBDT, a new boosting decision tree method with systematical uncertainties into training for High Energy Physics}
\author{Li-Gang Xia \\ Department of Physics, Warwick University, CV4 7AL, UK}

\begin{abstract}
    A new boosting decision tree (BDT) method, QBDT, is proposed for the classification problem in the field of high energy physics (HEP). In many HEP researches, great efforts are made to increase the signal significance with the presence of huge background and various systematical uncertainties. Why not develop a BDT method targeting the significance directly? Indeed, the significance plays a central role in this new method. It is used to split a node in building a tree and to
    be also the weight contributing to the BDT score. As the systematical uncertainties can be easily included in the significance calculation, this method is able to learn about reducing the effect of the systematical uncertainties via training. Taking the search of the rare radiative Higgs decay in proton-proton collisions $pp \to h + X \to \gamma\tau^+\tau^-+X$ as example, QBDT and the popular Gradient BDT (GradBDT) method are compared. QBDT is found to reduce the correlation between the signal strength and systematical uncertainty sources and thus to give a better significance. The contribution to the signal strength uncertainty from the systematical uncertainty sources using the new method is 50-85~\% of that using the GradBDT method.

\end{abstract}

\maketitle

\section{Introduction}
In the field of high energy physics (HEP), many machine-learning (ML) methods are used for object identification (ID)~\cite{atlas_tauid,atlas_photonid, atlas_bjetid,cms_tauid,cms_bjetid} and to search for rare signals~\cite{atlas_hww_emu, atlas_chargedhiggs, atlas_bbtautau, atlas_hbb, atlas_haa_4b, cms_hmumu, cms_hww, cms_hh_aabb, cms_haa, cms_htt}. Especially, artificial neutral network (ANN) and boosting decision tree (BDT) are two of the most popular methods. The latter one is found to be more robust and stable in some analyses~\cite{hjyang1, hjyang2}. On the other hand, various systematical uncertainties are involved in HEP (for examples, 13 main systematical uncertainties are considered in the analysis~\cite{atlas_hww_emu}) and they will inevitably affect the distribution of the input variables in training. However, there is no mature ML algorithm on training with systematical uncertainties as far as I know. 
This issue will be a big concern for either physicists or computer scientists in the near future as stated in Ref.~\cite{whitepaper}. 
Some efforts have been made. 
The pivot adversarial network (PAN) is proposed in Ref.~\cite{advNN} to make the score decision distribution robust against one systematical uncertainty by modifying the loss function. A case study in HEP following this method is performed in Ref.~\cite{victor}, but does not find significant improvement compared to the plain neural network. 
The figure of merit defined as the relative error of the signal strength in Ref.~\cite{victor} is stimulating and is actually closely related with the signal significance used in this work. 
In Ref.\cite{adam}, loss functions are designed to maximize the statistical significance in neural networks. This is a good direction, but it seems to treat systematical uncertainties in a loose way. It is unclear to me how this method and PAN could consider different systematical uncertainties and the correlations in a consistent way. 
The inference-aware neural network (INFERNO) proposed in Ref.~\cite{inferno} is to minimize the uncertainty for parameters of interest (the figure of merit mentioned above), obtained from the likelihood function, by updating the network output distribution. This method looks promising as it is convenient to incorporate various systematical uncertainties into the likelihood function, which idea is also used in this work. 
Ref.~\cite{inferno} shows the INFERNO method is better than the usual neural network from the perspective of signal strength uncertainty, but does not show the post-fit uncertainty of the nuisance parameters. This could be a concern if they are more over-constrained than the usual methods.

In this paper, we propose a new BDT method, QBDT, to improve the sensitivity of probing rare signal with the existence of systematical uncertainties. The codes for this algorithm can be found here~\cite{qbdt}. Intuitively, the method pays more attention to the parameter space (spanned by the input variables) with higher signal purity and smaller background uncertainties at the same time.  In HEP, the signficance is a quantity suitable to balance the different requirements on signal purity and background uncertainty. One idea is including
systematical uncertainties into the significance calculation, and using the significance to build decision trees and to act as (or at least part of) the BDT output. 

In this paper we only focus on the application of BDT methods in the classification problem in HEP. 
We will review current BDT methods in Sec.~\ref{sec:review_BDT} and present the QBDT algorithm in Sec.~\ref{sec:QBDT_algorithm}. Comparison of this method and the Gradient BDT will be performed in three successive examples described in Sec.~\ref{sec:ex1}, Sec.~\ref{sec:ex3} and Sec.~\ref{sec:ex7}, respectively. The conclusions will be summarized in Sec.~\ref{sec:summary}.

\section{Review of the BDT methods}~\label{sec:review_BDT}
In the first place, let us introduce some necessary concepts. For a typical classification problem in HEP, we have two categories, signal and background. We can assign them different values called truth value $Y$. It is 1 for signal events and -1 for background events by convention. Some observables are selected as input in the BDT training. They could be four-momenta of the final particles or various mass variables. They are denoted by a vector $\vx\equiv (x_1, x_2,\cdots)$.

There are two popular BDT methods. One is the Adaptive BDT (AdaBDT) and the other is the Gradient BDT (GradBDT). Both methods use decision trees as the weak learners. Usually, hundreds of trees are used and each tree contains several terminal nodes (denoted by $R$) which are regions defined with the input variables $\vx$. The terminal nodes do not overlap and one event will fall into one of them. Each tree provides an output, which is the classification result, and a tree weight, which reflects the probability of this classification being correct. The final output, called as BDT score $y(\vx)$, is a combination of the outputs from all trees. They are boosting algorithms because the subsequent decision tree will pay more attention to the events which are misclassified by previous trees. Let us call it ``tree update'' for convenience throughout this paper.  

For AdaBDT, the output of each tree is denoted by $k$, which is 1 if an instance is classified as a signal event (namely if it falls in a terminal node dominated by the signal events) and -1 if classified as a background event (namely if it falls in a terminal node dominated by the background events). The confidence of the classification is measured by a function of the misclassification rate. It is $\alpha \equiv \frac{1}{2}\ln\frac{1-\epsilon}{\epsilon}$ where $\epsilon$ is the misclassification rate for a tree. $\alpha$ is big if the misclassification rate is low. The final score is a combination, $y(\vx) = \sum_{i=1}^m k_i(\vx)\alpha_i$ for a $m$-tree training. In building a tree, the split of a node into two is determined by the Gini index . It is defined as $\sum_{\vx\in N}p(1-p)$ where $N$ denotes a node in a tree (keep in mind that we call it $R$ instead if it is a terminal node) and $p$ is the fraction of signal events (purity) in that node. The tree update in AdaBDT is realized by applying a big weight, $e^{\alpha}$, to those misclassified events and taking the weighted events as input in subsequent tree. It can be shown that AdaBDT is to minimize the loss function $L(y) = \sum_{\vx} e^{-y(\vx)Y(\vx)}$~\cite{friedman1,rojas}.

For GradBDT, we can use arbitrary differentiable loss function like $L(y) = \sum_{\vx} \frac{1}{2}(y(\vx)-Y(\vx))^2$ or $L(y) = \sum_{\vx}\frac{1}{1+e^{y(\vx)Y(\vx)}}$. GradBDT is to minimize the loss function in a stage-wise way. Let us describe the process in the following example. We start with a random guess, like 0 for all events. The first tree is trained to fit the difference between 0 and the truth values. The output of the first tree is denoted by $w_1(\vx)$. This is done by maximizing $L(y_0) - L(y_1)$ (thus the loss function will reduce after the first tree), where $y_1(\vx)=0+w_1(\vx)=w_1(\vx)$ for the first tree and $y_0(\vx)=0$ as the initial guess. It can be shown that $w_1(\vx)$ is roughly negative gradient~\cite{friedman2} of the loss function evaluated at $y_0=0$, $\frac{\partial L(y)}{\partial y}|_{y=y_0=0}$. The second tree is trained to fit the difference between $y_1(\vx)$ and the truth values. The output of the second tree is $w_2(\vx)$, which is also roughly negative gradient of the loss function evaluated at $y_1$. So $y_2(\vx)=y_1(\vx)+w_2(\vx)=w_1(\vx)+w_2(\vx)$. This tree update can be repeated and the final score $y_m(\vx) = \sum_{i=1}^m w_i(\vx)$ for a $m$-tree training. In GradBDT, the node split is naturally determined by minimizing the loss function. For example, a node $N$ may give two disjoint daughter nodes $N_L$ and $N_R$ depending on $x<x_0$ or $x>x_0$. The split position $x_0$ is determined by maximizing $\sum_{\vx \in N}l(y(\vx)) - \sum_{\vx \in N_L}l(y(\vx)) - \sum_{\vx \in N_R}l(y(\vx))$, where $l(y(\vx))$ is the loss function for one event. Recent development about GradBDT can be found in Ref.~\cite{xgboost}.

Table~\ref{tab:rev_bdt} compares the two methods as well as the new one as presented in next section. The first column lists the basic elements constituting a BDT algorithm. This table does not only show the features of different algorithms, but also helps to construct a new algorithm as long as all elements are defined reasonably.

\begin{table}
   \caption{\label{tab:rev_bdt} 
	Comparison of different BDT methods. Here $m$ denotes a $m$-tree training. $\epsilon$ is the misidentification rate in AdaBDT. $w$ is negative gradient of the loss function in GradBDT. $p$ and $Q$ is the purity and significance respectively of a terminal node in QBDT. 
   }
   \begin{tabular} {llll}
    \hline\hline
    Quantity &  AdaBDT value & GradBDT value & QBDT value\\
    \hline
    Input variables &$\vec{x}=(x_1,x_2,\cdots)$& same  & same\\
    True value $Y$ & -1, 1 & same  & same\\
    Tree output & $k=-1, 1$& negative gradient $w$  & $2p - 1$\\
    Tree weight & $\alpha = \frac{1}{2}\ln \frac{1-\epsilon}{\epsilon}$ & 1 & $\sum_{j=1}^J Q_j$ \\
    Tree update & apply $e^{\alpha}$ to wrong guess & fit the residues & apply $e^{\pm p\sum_j Q_j}$\\
    Node split & Gini index reduction & loss function reduction & significance $Q$ increasement \\
    BDT score $y_m$& $y_m = y_{m-1} + \alpha_m k_m $  & $y_m = y_{m-1} + w_m$ & $y_m=y_{m-1} + (2p-1)\sum_j Q_j$\\
    \hline
    Loss function $L_m(\vec{x}, y_m)$ & $\sum_{\vec{x}} e^{-Y(\vec{x})y_m(\vec{x})}$& any form & significance\\
    \hline\hline
 \end{tabular}
 \end{table}

\section{QBDT algorithm}~\label{sec:QBDT_algorithm}

The new BDT method is based on the concept of significance in HEP. For a measurement with the number of signal and background events being $s$ and $b$ respectively, the expected significance can be represented by the square root of the likelihood ratio denoted by $Q$ (this is why the new method is named as ``QBDT'')~\cite{asimov} with the definition
\begin{equation}\label{eq:Q0}
   Q \equiv 2\ln\frac{P(s+b|s+b)}{P(s+b|b)} = 2\left[ (s+b)\ln(1+\frac{s}{b}) - s\right]\:,
\end{equation}
where $P(n|\nu)\equiv\frac{\nu^n}{n!}e^{-\nu}$ is the Poisson probability distribution function (PDF) with the expectation value $\nu$. If $s<<b$ as in the case of searching for rare signals in HEP, we have $Q\approx \frac{s^2}{b}$. 

Let us explain the various elements for this algorithm with the help of Table~\ref{tab:rev_bdt}.

\begin{figure}
   \includegraphics[width=0.8\textwidth]{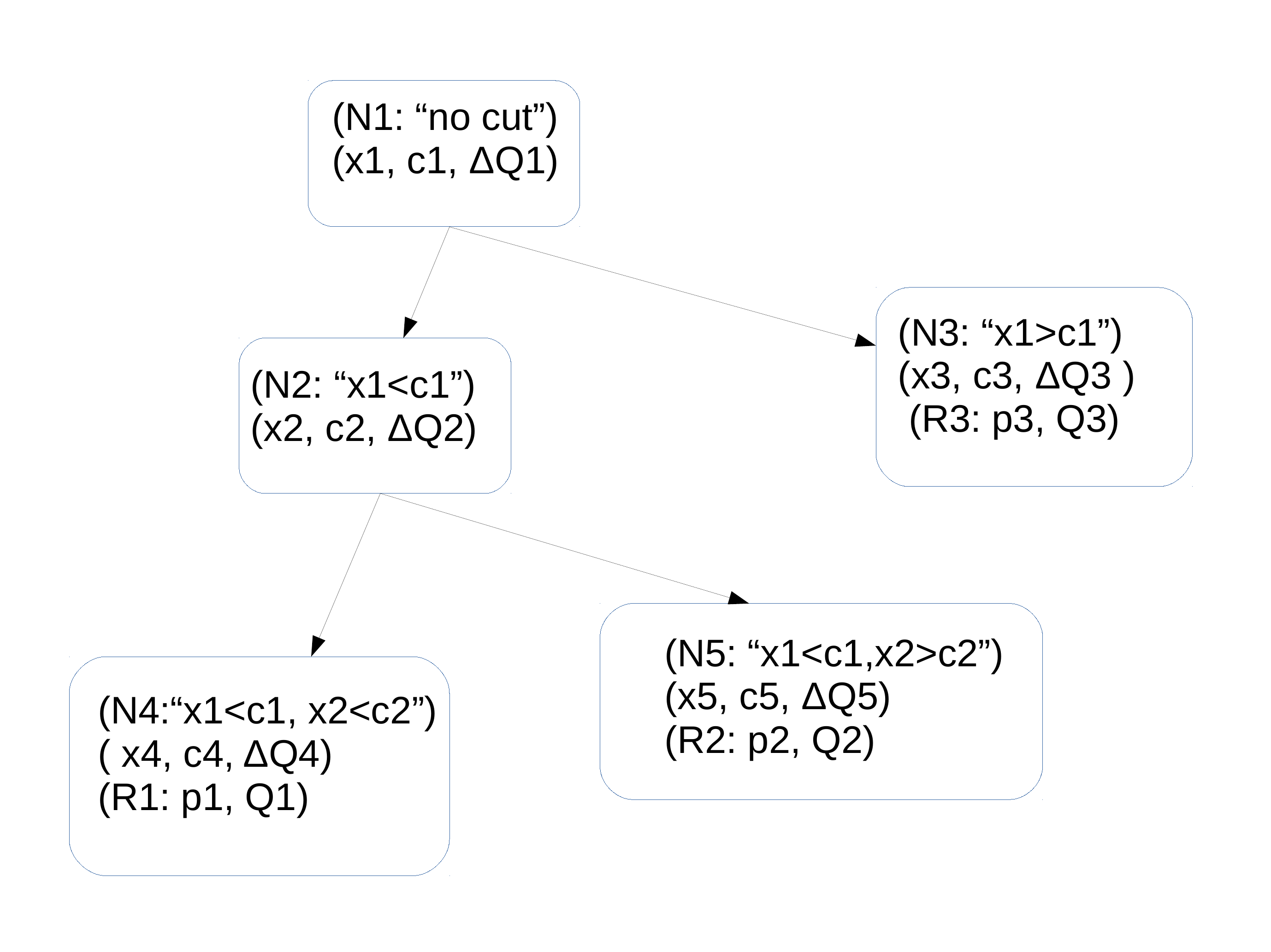}
   \caption{\label{fig:treesplit}
	Illustration diagram for the node splitting in QBDT method.
   }
\end{figure}

\begin{itemize}
   \item Input variables ($\vx$): same as other BDT methods
   \item True value: same as other BDT methods
   \item Tree output: this depends upon the terminal node. It is $2p_j-1$ if an event fall into the terminal node $R_j$. Here $p_j$ is the purity and defined as $p_j \equiv \frac{s_j}{s_j+b_j}$ with $s_j$ ($b_j$) being the number of signal (background) events in $R_j$. We use $2p-1$ as the output so that it is 1 for a region with purely signal events and -1 for a region with purely background events. Purity is the right variable to maximize the significance in the following sense. Suppose we have two regions $R_1$ and $R_2$, the total significance is approximately $\frac{s_1^2}{b_1} + \frac{s_2^2}{b_2}$ which is greater than or equal to the significance of merging two regions, namely, $\frac{s_1^2}{b_1} + \frac{s_2^2}{b_2} \geq \frac{(s_1+s_2)^2}{b_1+b_2}$. The equal sign holds if and only if $\frac{s_1}{b_1}= \frac{s_2}{b_2}$, which means two regions have the same purity. As each tree output is part of the BDT score and the score will be used as the observable to extract the signal strength, we should assign different scores to events falling into regions with different purities. Otherwise, the significance will not increase. 
   \item Tree weight: it measures the probability of correct classification based on the present tree. It is defined as the total significance in all terminal nodes, $\sum_{j=1}^J Q_j$. This is consistent with our intuitive understanding. The trees with higher total significance should contribute more to the final score. 
   \item Tree update: this is similar to AdaBDT. We apply a weight of $e^{p\sum_jQ_j}$ to the background events and a weight of $e^{-p\sum_jQ_j}$ to the signal events. It means that a background event happening to fall in a high-purity region or a signal event happening to fall in a low-purity region will be given a bigger weight in subsequent tree.
   \item Node split: this is similar to GradBDT. For any possible split about a variable $x$ with the split position $c$, let $Q$ be the significance of a node before splitting and $Q_L$ ($Q_R$) be the significance of the daughter node corresponding to $x<c$ ($x>c$). We search for the variable $x$ and the split position $c$ to maximize the significance increasement $\Delta Q(x,c)\equiv Q_L(x,c) + Q_R(x,c) - Q(x)$. 
	\begin{equation}
	   x,c = \arg\max_{x,c} \Delta Q(x,c)
	\end{equation}
	Figure~\ref{fig:treesplit} illustrates the growth of a tree. We starts with the whole set of events, denoted by ``$N_1$:no cut'' in the top. Performing the significance maximization procedure above gives the variable $x_1$ and the split position $c_1$. Then we have two nodes ``$N_2: x_1<c_1$'' and ``$N_3: x_1>c_1$''. For each of them, we repeat the significance maximization procedure to find the variable and split position, namely, $x_2$ and $c_2$ for $N_2$ with the significance increasement $\Delta Q_2$ and $x_3$ and $c_3$ for $N_3$ with the significance increasement $\Delta Q_3$. If $\Delta Q_2 > \Delta Q_3$, the next split will happen to $N_2$ and the daughter nodes are denoted by ``$N_4:x_1<c_1, x_2<c_2$'' and ``$N_5:x_1<c_1, x_2>c_2$''. If we require a tree to have only 3 terminal nodes, the splitting stops and the terminal nodes are $N_3$, $N_4$ and $N_5$, which are renamed as $R_1$, $R_2$ and $R_3$ in Fig.~\ref{fig:treesplit}. For each of them, we can calculate the purity $p_j$ and the significance $Q_j$ ($j=1,2,3$). $2p_j-1$ is the output of the tree while $\sum_j Q_j$ is the tree weight. If we allow more terminal nodes, the next split will happen to one of $N_4$, $N_5$ and $N_3$, which gives the largest $\Delta Q$. Note that for each split, we allow to use the same variable, like $x_2 = x_1$.
   \item BDT score: the definition is shown below
	\begin{equation}~\label{eq:qbdt_score}
	y_m(\vx)= \sum_{i=1}^m \sum_{j=1}^J \delta_{\vx, R_j^{(i)}}(2p_j^{(i)}-1) \sum_{k=1}^J Q_k^{(i)}
   \end{equation}
   for $m$ trees, where 
   $p_j^{(i)}$ is the purity of the terminal node $R_j^{(i)}$ in the $i$-th tree; 
	$Q_k^{(i)}$ is the significance of the terminal node $R_k^{(i)}$ in the $i$-th tree; 
	and $\delta_{\vx,R_j^{(i)}} =1$ if $\vx \in R_j^{(i)}$ and 0 otherwise.
\end{itemize}

 The algorithm of QBDT should be clear after all elements are defined. Here is the workflow.
 \begin{enumerate}
    \item Reweight both signal and background samples so that each gives 1 event. This reweighting is not necessary, but very helpful in practice. It is because the significance is tiny for a rare signal (but this is usually where BDT is used) and the boosting is very slow. Many more trees would be needed without reweighting.
    \item According to the node split procedure, we can build the first tree. Supposing only $J$ terminal nodes are allowed, we can calculate the purity $p_j^{(1)}$ and the significance $Q_j^{(1)}$ for each terminal node $R_j^{(1)}$. The tree weight is the total significance, $\sum_{j=1}^J Q_j^{(1)}$. If an event falls into $R_x^{(1)}$, then the output is $(2p_x^{(1)}-1)\sum_{j=1}^J Q_j^{(1)}$.
    \item Before building the second tree, we apply a weight of $\exp(p_{j_b}^{(1)}\sum_{j=1}^J Q_j^{(1)})$ to every background event (supposing it falls into $R_{j_b}^{(1)}$) and similarly apply a weight of $\exp(-p_{j_s}^{(1)}\sum_{j=1}^J Q_j^{(1)})$ to every signal event (supposing it falls into $R_{j_s}^{(1)}$). Reweight both signal and background samples to 1 event again.
    \item Build the second tree based on the reweighed samples using the same node split procedure. We have $J$ terminal nodes, $R_j^{(2)}$ and the corresponding purity $p_{j}^{(2)}$ and significance $Q_j^{(2)}$. If an event falls into $R_x^{(2)}$, then the output is $(2p_x^{(2)}-1)\sum_{j=1}^J Q_j^{(2)}$.
    \item We can repeat the steps 2--4 to build many more trees. The final output (BDT score) is given by Eq.~\ref{eq:qbdt_score}.
 \end{enumerate}

 The key feature of QBDT is that it is able to consider systematical uncertainties into training. In HEP, various factors will affect the input variables. For example, if the muon transverse momentum is one input variable, we have to evaluate the uncertainty due to muon momentum calibration. Furthermore, the identification efficiency and isolation efficiency of the muon lepton may be dependent upon the transverse momentum. Thus the efficiency uncertainties will also affect the transverse momentum distribution. To include these uncertainties into training, the definition of $Q$ can be extended to be~\cite{glen_Qsyst}
 \begin{equation}\label{eq:Q}
    Q \equiv 2\left[ (s+b)\ln\frac{(s+b)(b+\sigma_b^2)}{b^2+(s+b)\sigma_b^2}-\frac{b^2}{\sigma_b^2}\ln[1+\frac{s\sigma_b^2}{b(b+\sigma_b^2)}] \right] \:,
 \end{equation}
 where $\sigma_b$ is the systematical uncertainty on the background number of events. If $s<<b$ and $\sigma_b<<b$, $Q\approx \frac{s^2}{b+\sigma_b^2}$. The purity in the tree output is then extended to be $\frac{s}{s+b+\sigma_b^2}$ accordingly. 

 In the case of multiple systematical uncertainties, we may need to consider the possible correlations. The total background uncertainty will be 
 \begin{equation}~\label{eq:sigmab0}
    \sigma_b^2 = \sum_{i=1}^{\Nsysts} \left(\frac{\partial b}{\partial \theta_i}\sigma_{\theta_i}\right)^2 + \sum_{i\neq j} \frac{\partial b}{\partial \theta_i} \frac{\partial b}{\partial \theta_j} \sigma_{\theta_i}\sigma_{\theta_j} \rho_{ij} \:, 
 \end{equation}
where each $\theta_i$ is a nuisance parameter associated with one systematical uncertainty source; $\sigma_{\theta_i}$ is the size of the uncertainty which is usually estimated by other independent measurements; $\frac{\partial b}{\partial \theta_i}\sigma_{\theta_i}$ is the size of the background number uncertainty due to this systematical uncertainty source; $\rho_{ij}$ is the correlation coefficient between two systematical uncertainty sources. 
In most of HEP cases, we perform a likelihood fit to the distribution of the final BDT score. The likelihood function consider all systematical uncertainties by introducing nuisance parameters and can also consider the prior constraint on the nuisance parameters. 
For a binned BDT distribution with $\Nbins$ bins, the likelihood function is
\begin{equation}\label{eq:likelihood}
    \mathcal{L}(\mu,\theta_1,\cdots,\theta_{\Nsysts}) = \Pi_{i=1}^{\Nbins}P(n_i|\mu s_i+b_i+\sum_{j=1}^{\Nsysts}\theta_j\Delta_i^{j})\times\Pi_{j=1}^{\Nsysts}G(\theta_j|0,1) \:,
\end{equation}
where the index $i$ denotes the $i$-th bin and the index $j$ denotes the $j$-th systematical uncertainty source; $n_i$, $s_i$ and $b_i$ denote the number of data events, signal events, and background events in the $i$-th bin, respectively; $\mu$ is the parameter of interest (signal strength) and $\theta_j$s are nuisance parameters to take into account the systematical uncertainty $\Delta_i^j$; $P(n|\lambda)$ is the Poisson distribution function with the expectation $\lambda$;
and $G(\theta_j|0,1)\equiv \frac{1}{\sqrt{2}}e^{-\frac{\theta_j^2}{2}}$ is a normal distribution to introduce the prior constraint on $\theta_j$ (with its standard deviation renormalized to be 1 by construction). 
The exact uncertainty size and correlation can only be obtained after a fit (maximizing the likelihood function and evaluating the Hessian matrix). So $\sigma_b^2$ cannot be calculated according to Eq.~\ref{eq:sigmab0} without the fitting result.
Fortunately, explicit expression of the uncertainty on the signal strength, $\sigma_s$ , is derived in Ref.~\cite{constraint_xia} assuming small correlations in the maximum likelihood estimation method. Noting that $\sigma_s^2 = b + \sigma_b^2 + \delta^2$ (Eq.(6) in Ref.~\cite{constraint_xia}) where $\delta$ is the Monte Carlo (MC) statistical uncertainty (it exists if any background component is estimated by MC simulation) and using the expression on $\sigma_s$ (Eq.(26) in Ref.~\cite{constraint_xia}), we have
 \begin{equation}\label{eq:sigmab}
     \sigma_b^2 = b\left(\sum_{j=1}^{\Nsysts} \frac{(\Delta^j)^2}{b+(\Delta^j)^2} - \sum_{i\neq j} \frac{(\Delta^i)^2(\Delta^j)^2}{(b+(\Delta^i)^2)(b+(\Delta^j)^2)}\right) \: .
 \end{equation}
 Here $\Delta^i$ is the uncertainty on the background number of events due to the $i$-th systematical uncertainty source; and the second term in the big brackets represents the contribution due to the correlation between different systematical uncertainty sources. It deserves mentioning that we can consider MC statistical uncertainty in training as well by replacing $\sigma_b^2$ by $\sigma_b^2 + \delta^2$. 
 We can also see that $Q \approx \frac{s^2}{\sigma_s^2}$, which is the figure of merit used in Ref.~\cite{victor} and the optimization goal in the INFERNO algorithm~\cite{inferno}.

It is worth detailing how we obtain $\sigma_b^2$ for each terminal node. Every node is defined by a set of selection conditions and all nodes do not overlap. Taking $R_1$ in Fig.~\ref{fig:treesplit} as example, it is defined as ``$x_1<c_1, x_2<c_2$''. Let $b$ be the number of events, passing the condition ``$x_1<c_1,x_2<c_2$'', from the nominal background sample. 
Various systematical uncertainty sources would affect the reconstruction of the variables $x_1$ and $x_2$ (including the differential distribution and acceptance) and thus have different predictions. Let $b_i$ be the number of background events passing the same conditions corresponding to the $i$-th systematical uncertainty. Then $\Delta^i$ in Eq.~\ref{eq:sigmab} is defined as $b_i-b$.
Looping over all systematical uncertainty sources, we can calculate $\sigma_b^2$ for every node according to Eq.~\ref{eq:sigmab}.
In practice, $\sigma_b^2$ is found to be negative sometimes. There could be two reasons. One is that the correlation term has very large negative contribution and the expression is not valid as small correlation is assumed in the derivation in Ref.~\cite{constraint_xia}. The other is that large fluctuation would happen in some nodes due to limited training samples. We set it to $0$ in this case. 

 In the following sections, we will compare the performance of GradBDT and QBDT based on an HEP example. We do not use the BDT score directly, but a monotonic function about it. The GradBDT algorithm in the package TMVA~\cite{tmva} uses $\tanh(y)$ as the final score. The advantage of this function is that it maps $(-\infty, +\infty)$ to a bounded region $(-1, 1)$.  We will use a similar form $\tanh(cy)$, where $c$ is a constant and is optimized to be $0.65$ for GradBDT ($0.65$ is better than 1 used in the TMVA package at least for the example presented below) while $0.25$ for QBDT.
 We will perform likelihood fits to binned BDT score distributions with 20 bins from -1 to 1. $c$ is scanned from 0.05 to 1.5 with a step of 0.05 and the best value is found when $Q$ in Eq.~\ref{eq:Q0}, extended to a binned distribution, is the greatest.

\section{One example to compare GradBDT and QBDT}~\label{sec:example}
\subsection{Description of the example}~\label{sec:ex0}
\begin{figure}
   \includegraphics[width=0.22\textwidth]{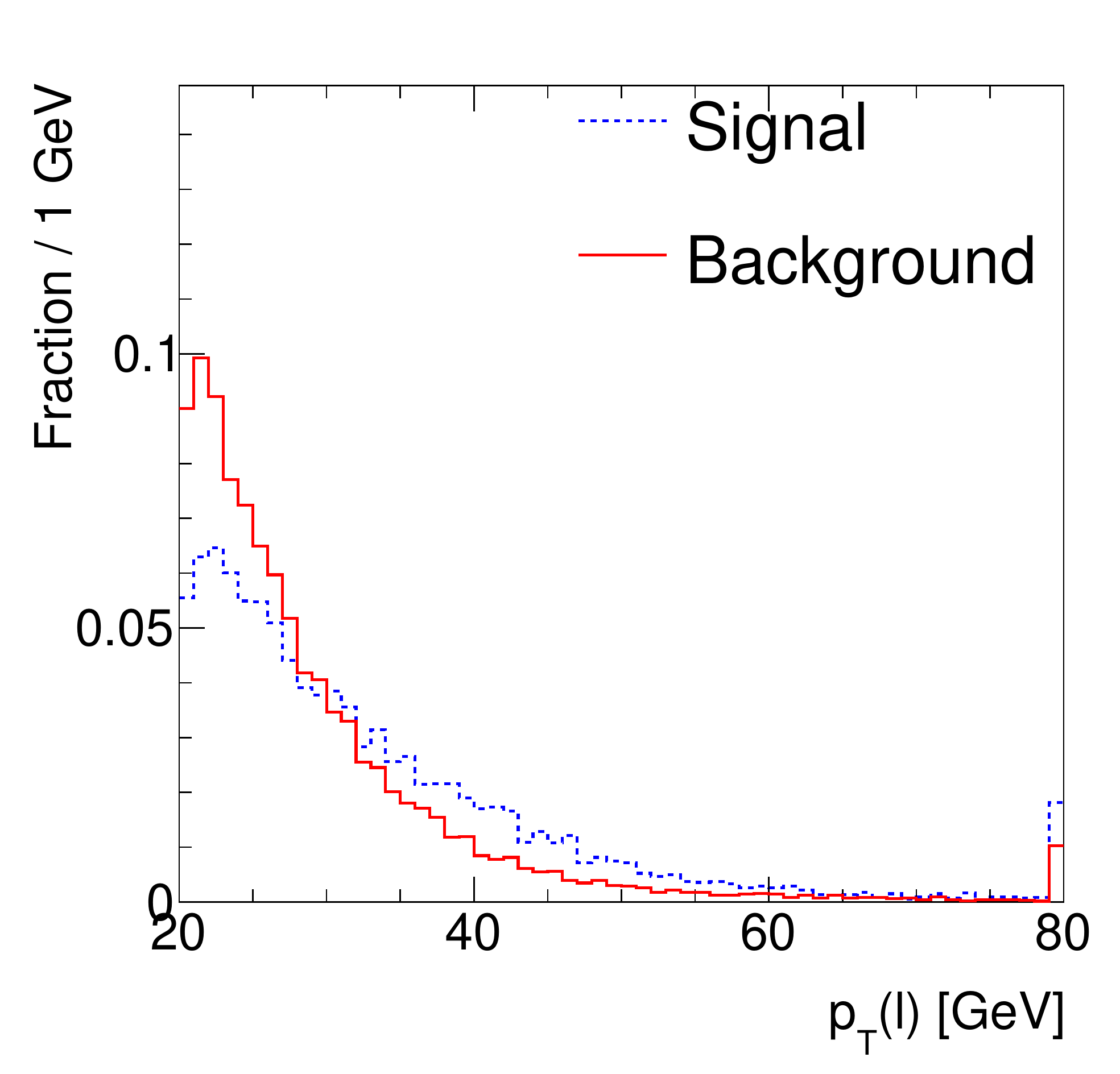}
   \includegraphics[width=0.22\textwidth]{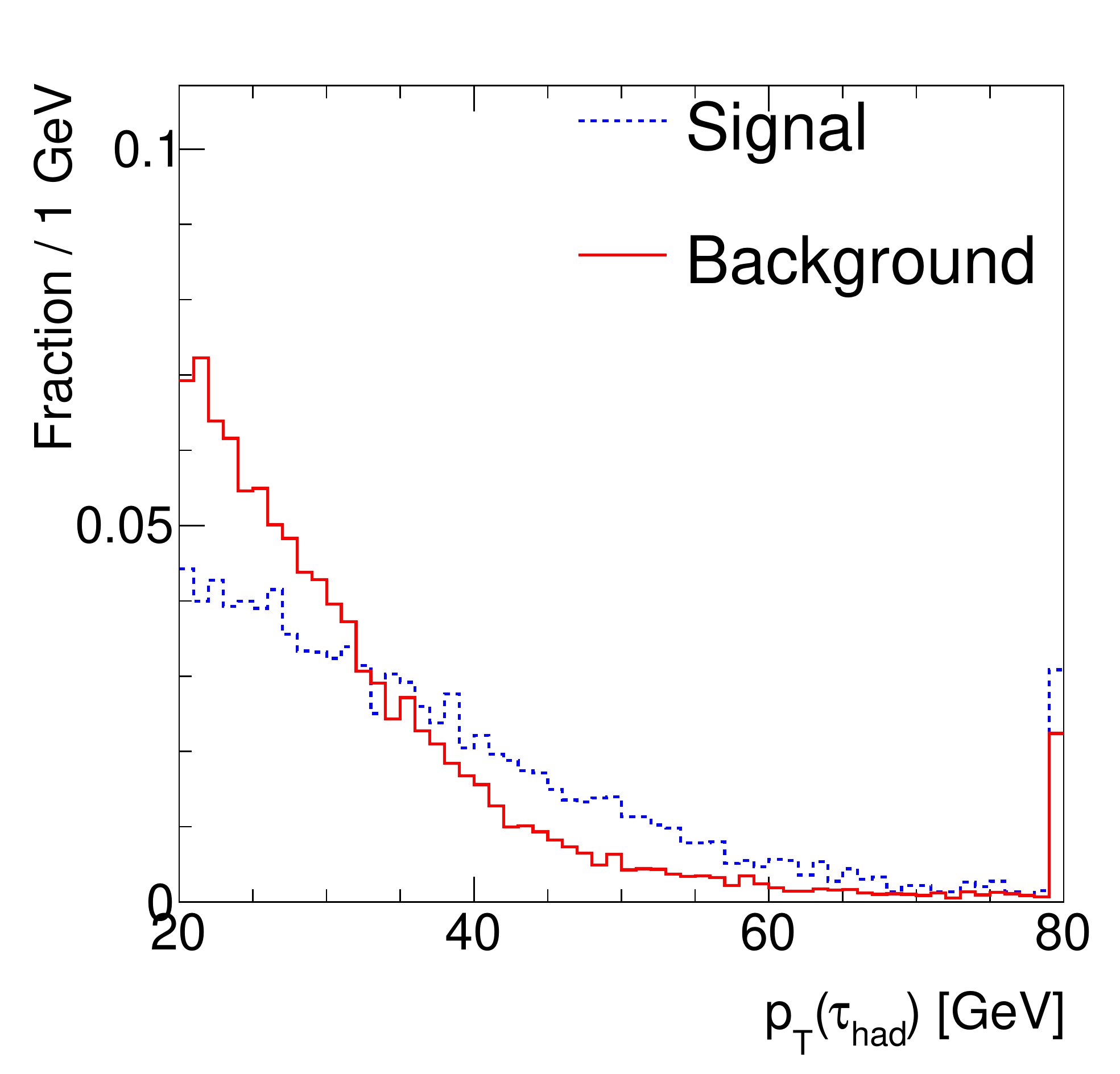}
   \includegraphics[width=0.22\textwidth]{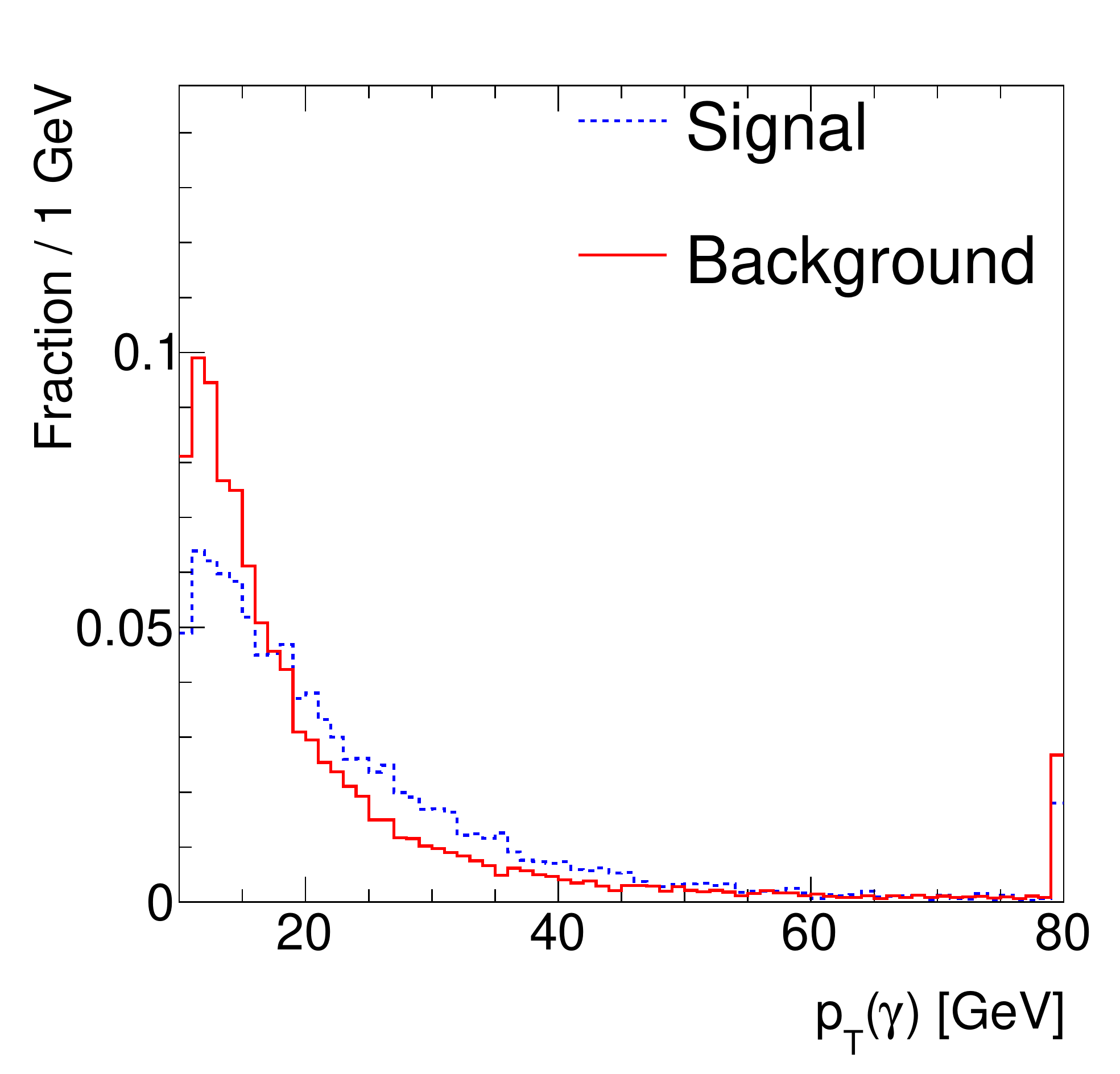}
   \includegraphics[width=0.22\textwidth]{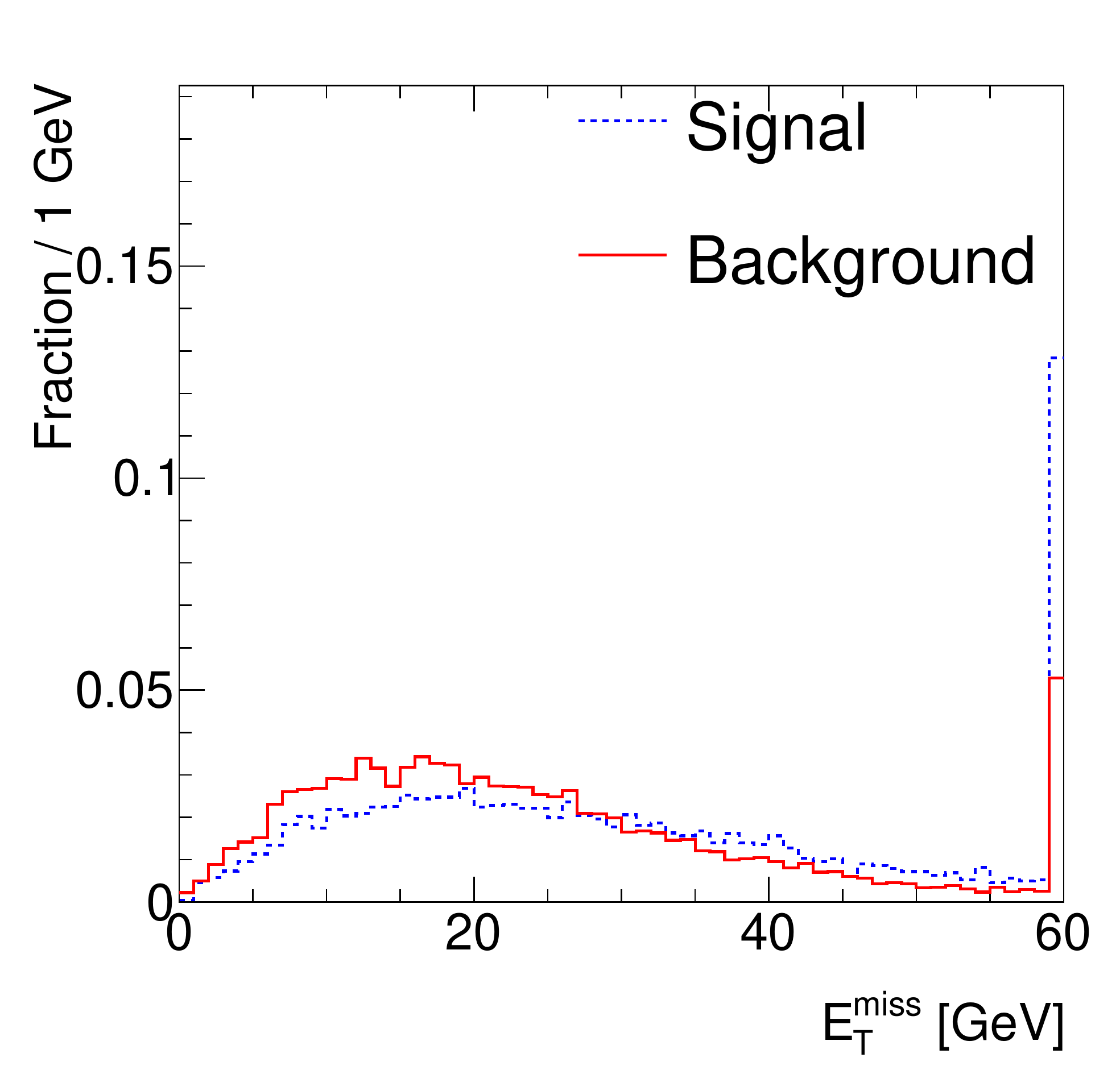}\\
   \includegraphics[width=0.22\textwidth]{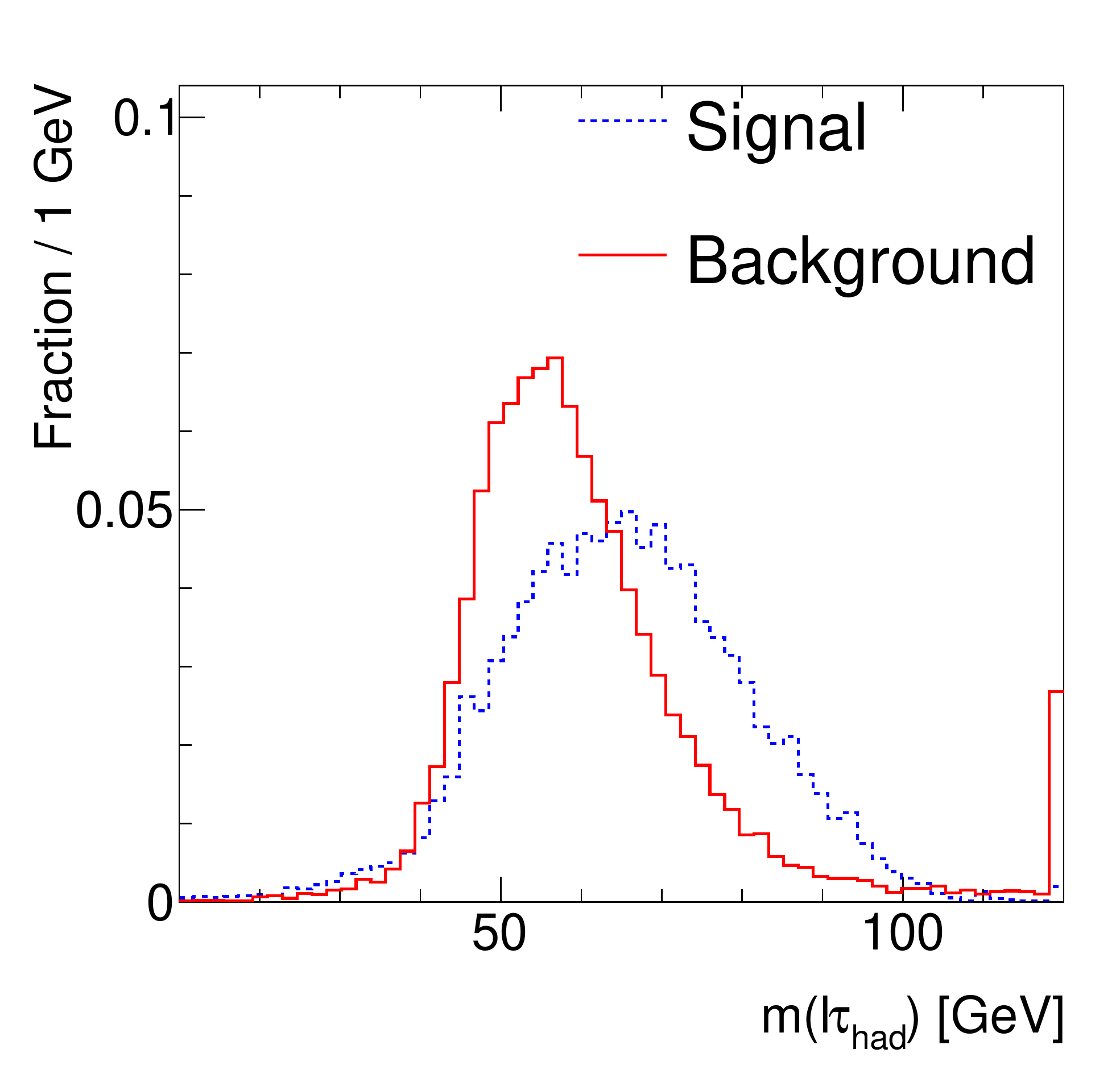}
   \includegraphics[width=0.22\textwidth]{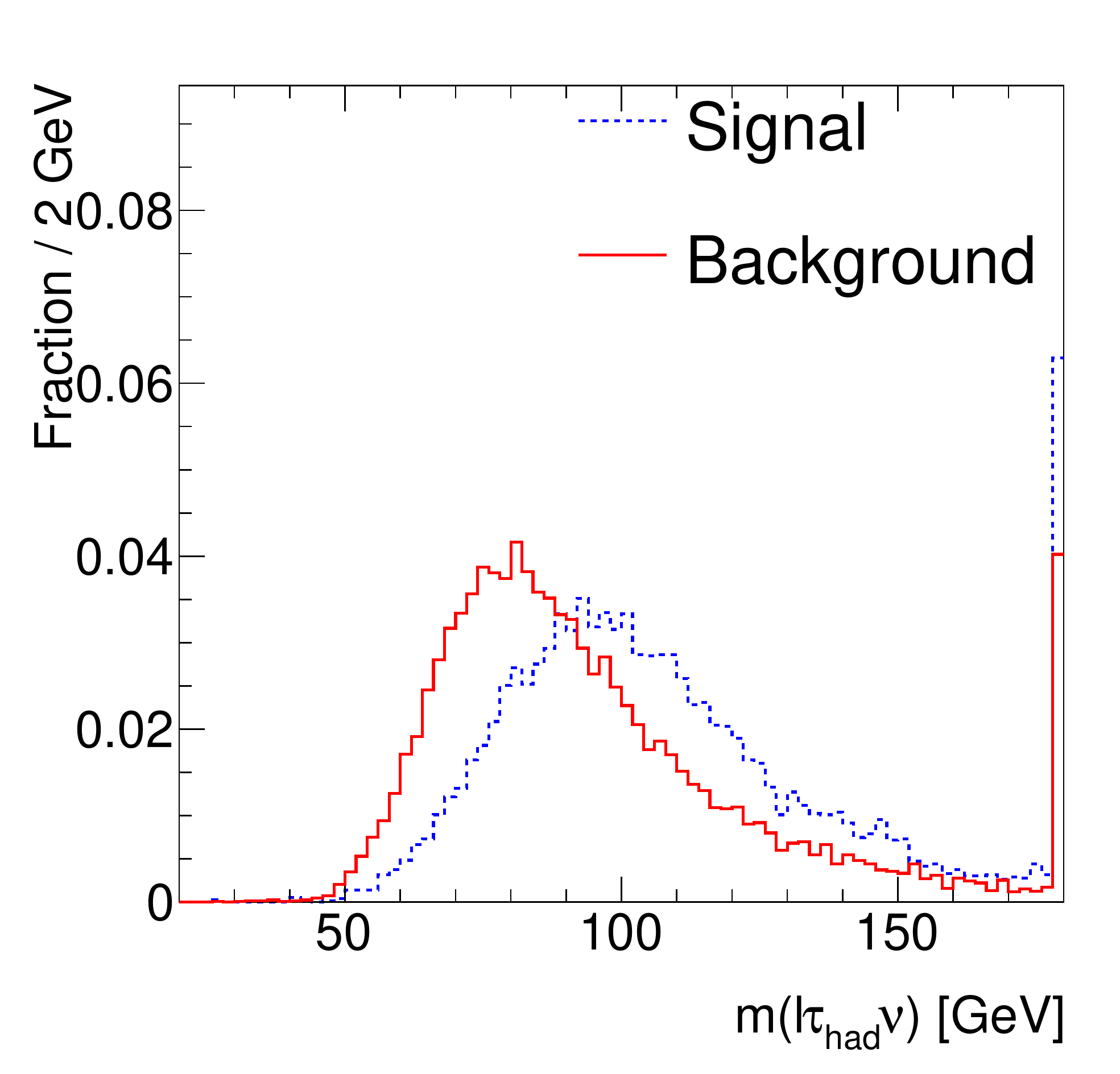}
   \includegraphics[width=0.22\textwidth]{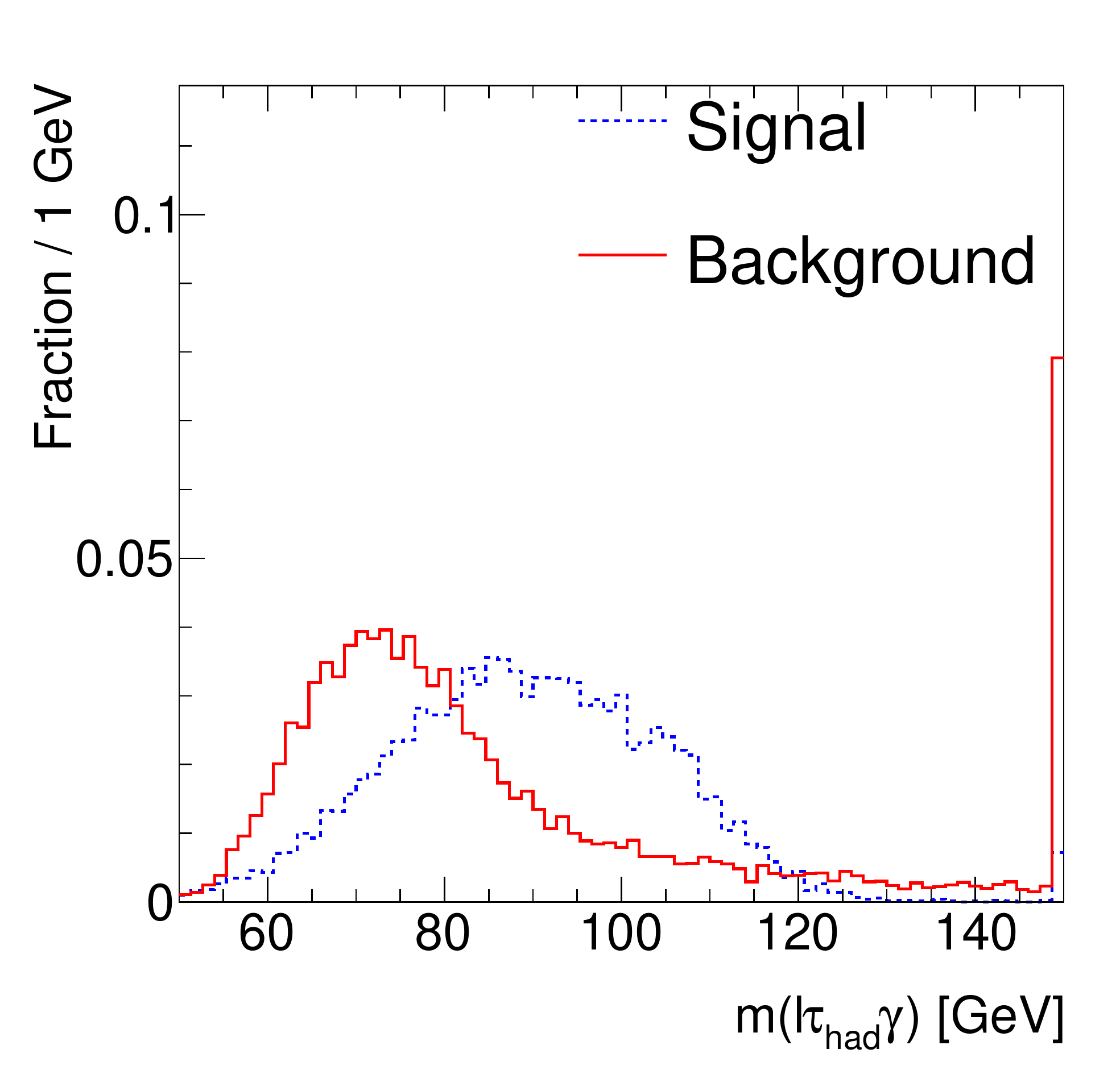}
   \includegraphics[width=0.22\textwidth]{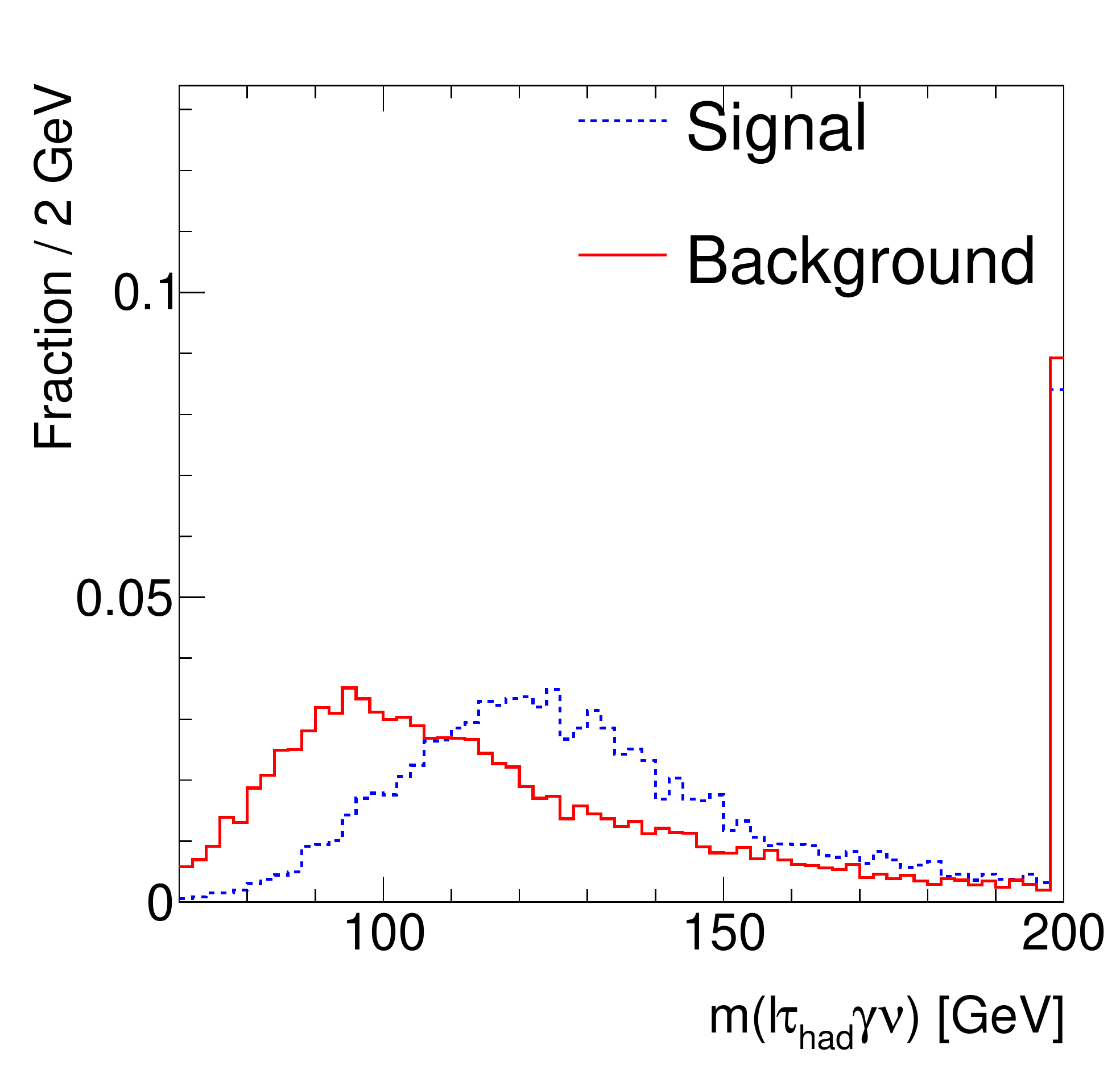}
   \caption{\label{fig:ex1_vars}
   (color online) The distribution of eight variables used in the training. In top row, it is $\pT(l)$, $\pT(\tauhad)$, $\pT(\gamma)$ and $\etmiss$ from left to right. In the bottom row, it is $m(l\tauhad)$, $m(l\tauhad\nu)$, $m(l\tauhad\gamma)$ and $m(l\tauhad\gamma\nu)$ from left to right. The blue dashed histogram represents the signal while the red solid histogram represents the background. All distributions are renormalized to unit area.  }
\end{figure}

Here we present an example. It is to search for the rare radiative decay of the Higgs boson in the proton-proton collisions at the center-of-mass energy $\sqrt{s}=13$~TeV assuming a dataset of 80~fb$^{-1}$. The signal process is $pp \to h+X \to \gamma\tau\tau+X$, while the background process is $pp \to \gamma\tau\tau + X$ with the Higgs contribution vetoed. 
We further require one tau decays leptonically, $\tau^- \to l^-\nu_\tau\bar{\nu}_\tau+ c.c.$ ( here $l$ is $e/\mu$ and $c.c.$ means the charge-conjugated channel), and the other decays hadronically, $\tau^- \to h^-\nu_\tau+c.c.$. 
The MC events are generated using MadGraph~\cite{madgraph} at leading order and the showering is simulated with Pythia8~\cite{pythia8, pythia6}. We did not use Delphes~\cite{delphes} to simulate the detector response, but the resolution effect is considered for the transverse momentum of final objects and the transverse missing energy by Gaussian smearing according to the performance of the ATLAS detector (see references in Table~\ref{tab:ex1_systs}). This should be enough for a toy example. 
The events with one charged lepton candidate $e/\mu$, one hadronic tau candidate $\tauhad$ and one photon candidate $\gamma$ are selected. The transverse momentum, $\pT$, is required to be greater than 20~GeV for the lepton and tau candidates and 10~GeV for the photon candidate. The missing transverse energy $\etmiss$ is defined as the negative vector sum of transverse momenta of all other visible objects. It reflects the information of the neutrinos.
Eight variables are used in the training.
Table~\ref{tab:ex1_variables} summarizes the variables.
They are the transverse momentum/energy of the final objects ( $\pT(l)$, $\pT(\tauhad)$, $\pT(\gamma)$ and $\etmiss$) and four mass variables constructed from the four-momenta (denoted by $p^4=(E,p_x, p_y,p_z)$) of the final objects. For the neutrinos, the transverse momentum is approximately $\etmiss$ while $z$ component cannot be known. Here we use $p^4(\nu) = (\etmiss, E_{\text{T,x}}^{\text{miss}}, E_{\text{T,y}}^{\text{miss}}, 0)$. The signal and background distributions of all variables are shown in Fig.~\ref{fig:ex1_vars}.

\begin{table}
   \caption{\label{tab:ex1_variables} 
   Definition of the eight variables used in the training.
   }
   \begin{ruledtabular}
	\begin{tabular}{l | l}
	   Variable & Explanation \\
	   \hline
	   $\pT(l)$ & $\pT$ of the charged lepton $e/\mu$ \\
	   $\pT(\tauhad)$ & $\pT$ of the hadronic tau \\
	   $\pT(\gamma)$ & $\pT$ of the photon \\
	   $\etmiss$ & missing transverse momentum \\
	   $m(l\tauhad)$ & invariant mass of the lepton $e/\mu$ and the hadronic tau \\
	   $m(l\tauhad\nu)$ & invariant mass of the lepton $e/\mu$, hadronic tau and neutrinos \\
	   $m(l\tauhad\gamma)$ & invariant mass of the lepton $e/\mu$, hadronic tau and photon \\
	   $m(l\tauhad\gamma\nu)$ & invariant mass of the lepton $e/\mu$, hadronic tau, photon and neutrinos \\
	\end{tabular}
   \end{ruledtabular}
\end{table}

For both GradBDT and QBDT, half of the samples are used for training and the other half used for testing. For GradBDT, we are using 1000 trees, learning rate of 0.1 and maximal depth of 3. These numbers are optimized.
For QBDT, we are using 100--200 trees and 7 terminal nodes for each tree. Increasing either the number of trees or the number of terminal nodes leads to little improvement. We have four setups for QBDT training. 

\begin{itemize}
   \item QBDT0: training with nominal samples, i.e., no systematical uncertainty considered in the training. 100 trees are used in training.
   \item QBDTX: $X=1,3,7$ is an integer denoting the number of systematical uncertainty sources considered in the training. 100 trees are used in training for $X=1,3$ and 200 trees for $X=7$.
\end{itemize}

Different QBDT setups as well as the GradBDT are compared in three cases, namely, the case of one systematical uncertainty source in Sec.~\ref{sec:ex1}, three sources in Sec.~\ref{sec:ex3} and seven sources in Sec.~\ref{sec:ex7}.
Table~\ref{tab:ex1_systs} summarizes all systematical uncertainties considered in this paper.
Here the uncertainty sizes are selected according to the performance~\cite{tau_perf,met_perf,muon_perf,photon_perf} of the ATLAS detector at the Large Hadron Collider (LHC).

\begin{table}
   \caption{\label{tab:ex1_systs} 
   Summary of systematical uncertainty sources.
   }
   \begin{ruledtabular}
	\begin{tabular}{l | l l}
	   systematical source  &Uncertainty size & Nuisance parameter\\ 
	   \hline
	   $\tauhad$ $\pT$ calibration  &$\Delta \pT(\tauhad) = \pm 2\% \pT(\tauhad)$~\cite{tau_perf} &  $\theta(\tau \pT)$\\
	   $\tauhad$ ID efficiency  & $\pm5\% $~\cite{tau_perf}   & $\theta(\tau\ID)$ \\
	   $\MET$ resolution & $\pm 10\%$~\cite{met_perf} & $\theta(\MET)$\\
	   Lepton $\pT$ calibration & $\Delta \pT(l) = \pm 0.1\% \pT(l)$~\cite{muon_perf} & $\theta(l\pT)$ \\
	   Lepton ID efficiency & $\pm 0.2\%$ for $\pT(l)<100$~GeV and $\pm 0.5\%$ otherwise~\cite{muon_perf} & $\theta(l\ID)$ \\
	   Photon $\pT$ calibration & $\Delta \pT(\gamma) = \pm 0.3\% \pT(\gamma)$~\cite{photon_perf} & $\theta(\gamma \pT)$ \\
	   Photon ID efficiency & $\pm2\%$ for $\pT(\gamma)<40$~GeV and $\pm1\%$ otherwise~\cite{photon_perf} & $\theta(\gamma\ID)$ \\
	\end{tabular}
   \end{ruledtabular}
\end{table}

\begin{figure}
   \includegraphics[width=0.6\textwidth]{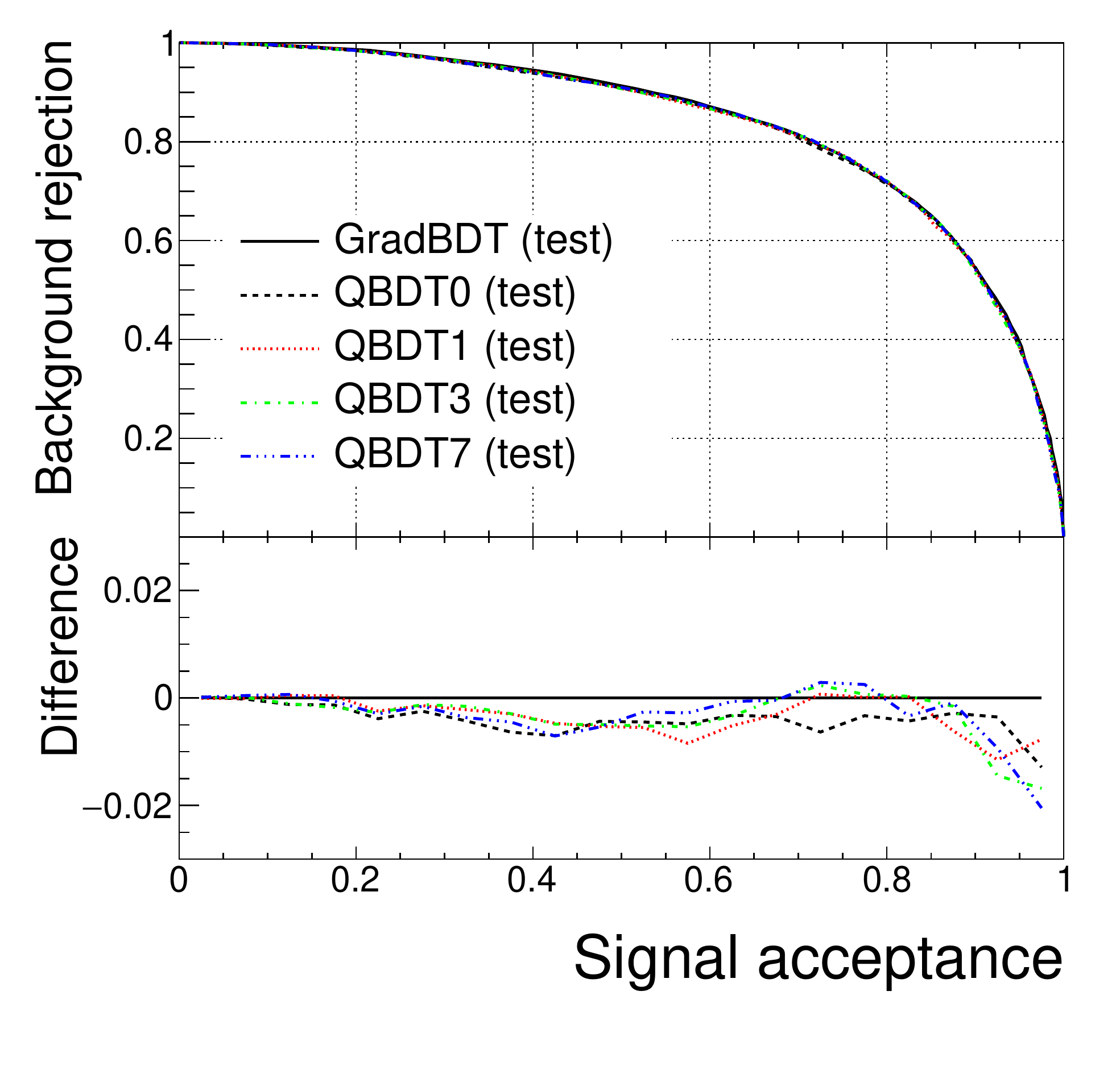}
   \caption{\label{fig:all_rocs}
       (color online) ROC curves from the testing samples. They are very close to each other. The lower pad shows the difference with respective to the GradBDT curve.
   }
\end{figure}

To get an overall impression about the performance of different BDT setups, we can look at the so-called ROC curves. A ROC curve describes the background rejection rate as a function of the signal acceptance if we apply a cut on the BDT score. Figure~\ref{fig:all_rocs} shows the ROC curves for all BDT setups. 
They are actually very close to each other with a difference of about $1\%$ for signal acceptance of $90\%$ as shown in the lower pad of Fig.~\ref{fig:all_rocs}. It should be mentioned that GradBDT seems better for most of the signal acceptance range. But ROC curve cannot fully measure the goodness of the differential distribution concerning the signal-background separation. All comparisons will be based on the testing samples in the following sections.

\subsection{Case I: one systematical uncertainty source}~\label{sec:ex1}

\begin{figure}
   \includegraphics[width=0.32\textwidth]{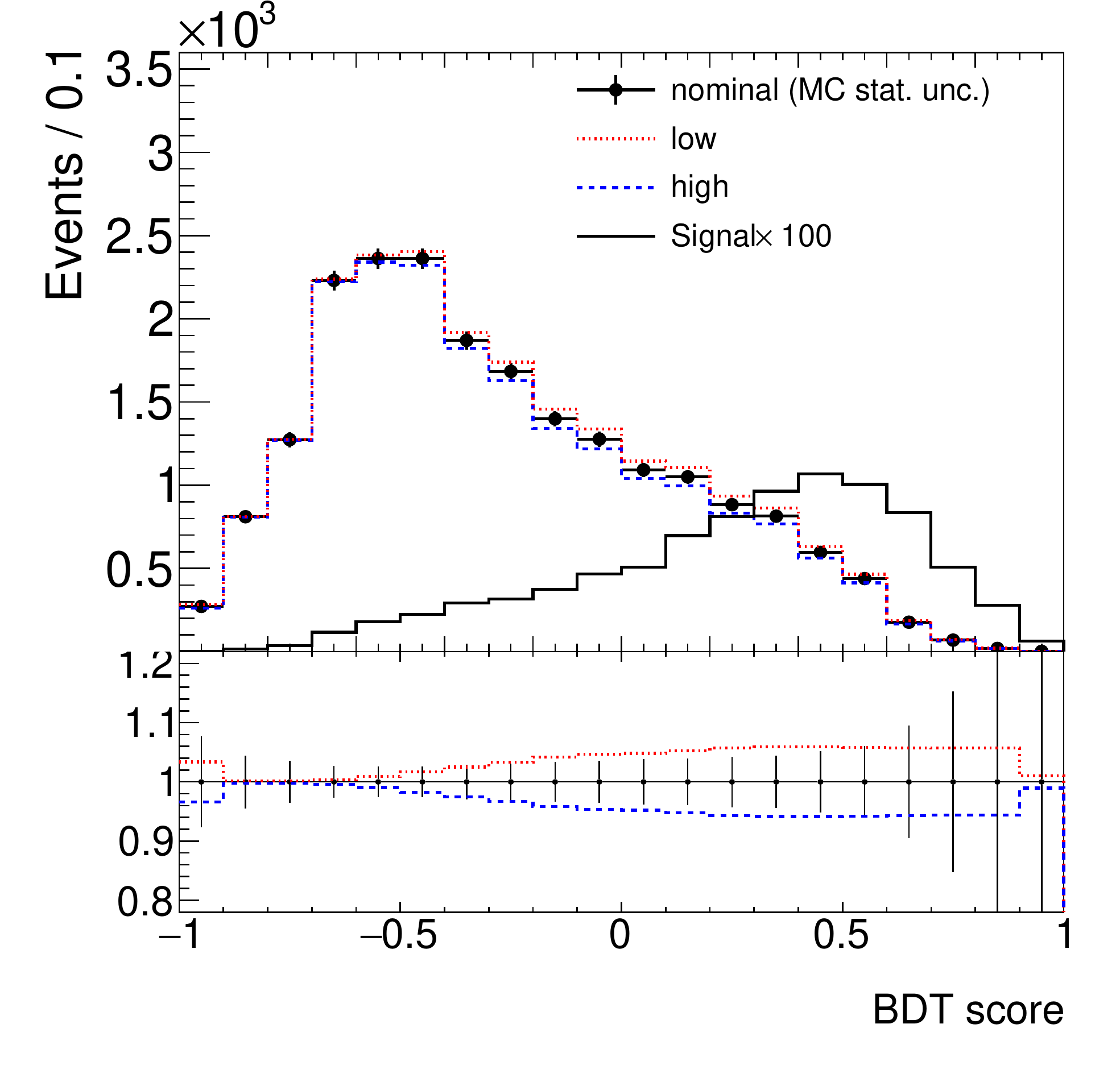}
   \includegraphics[width=0.32\textwidth]{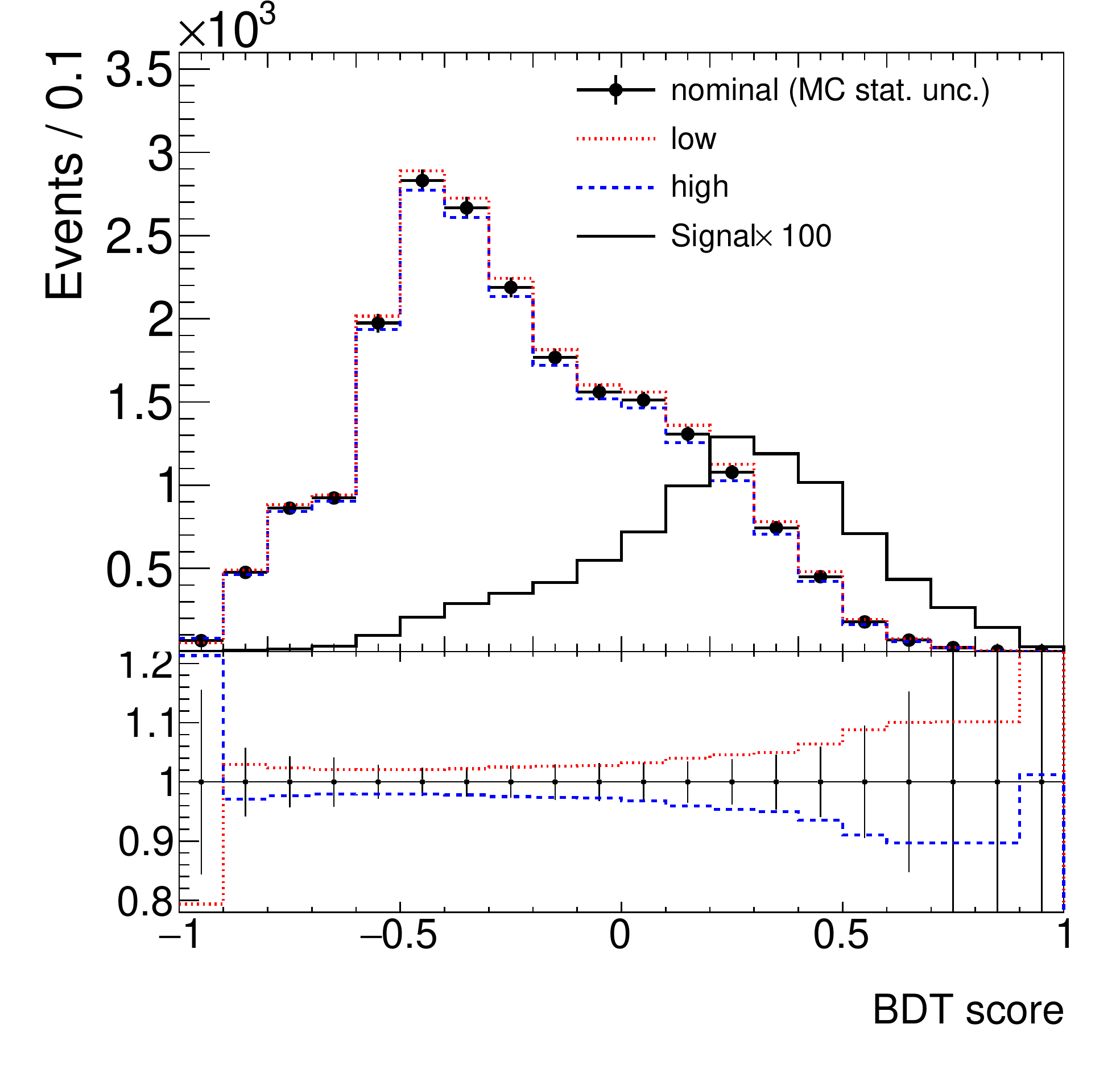}
   \includegraphics[width=0.32\textwidth]{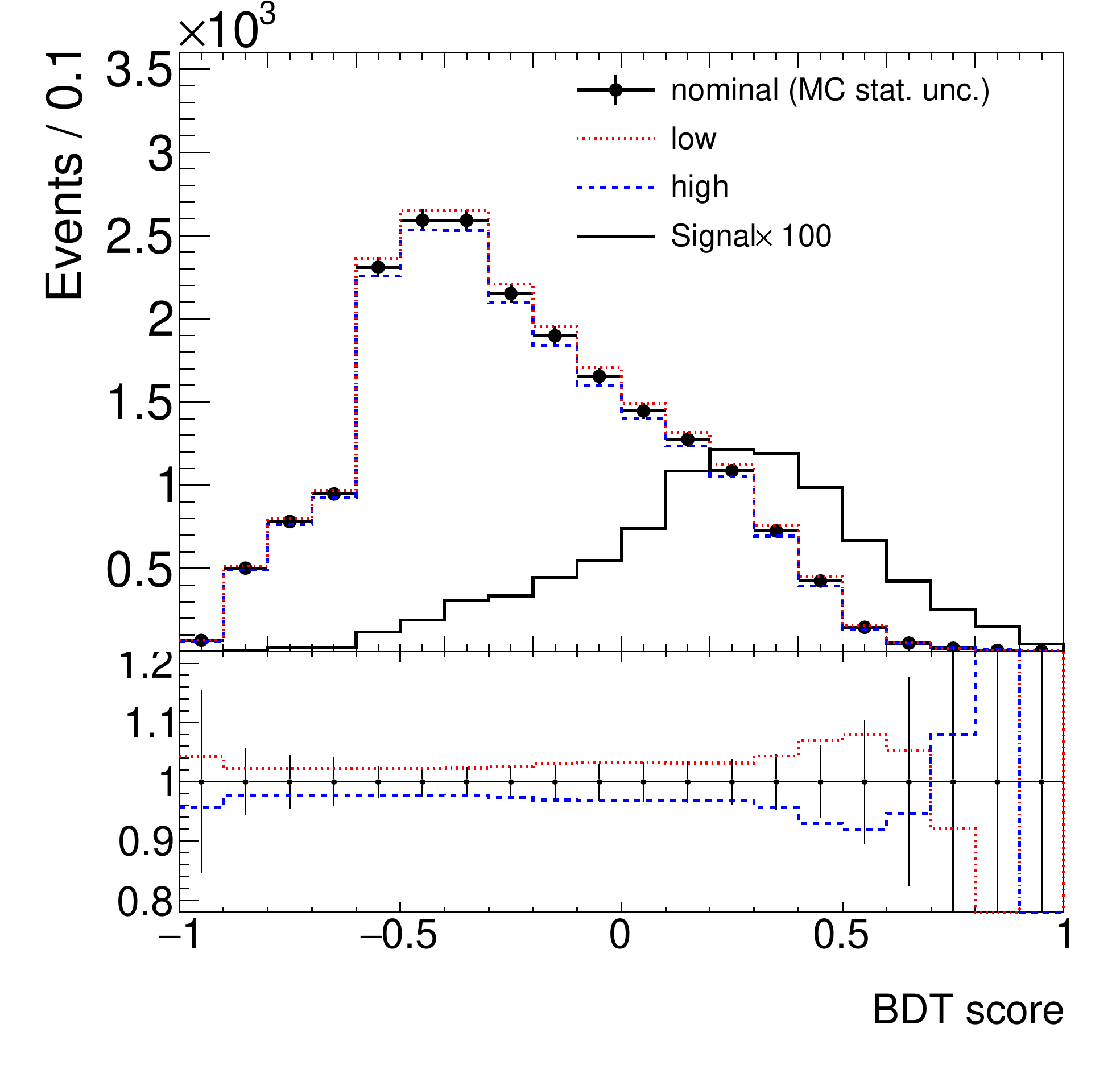}
   \caption{\label{fig:ex1_envelope_tes}
       (color online) Envelope plot for the systematical uncertainty of $\pT(\tauhad)$. Left: GradBDT, middle: QBDT0, and right: QBDT1. 
       The black dots with error bar represent the background. The plain black histogram represents the signal (scaled by a factor of 100 for illustration). The blue dashed/red dotted histograms represent the high/low variation of a systematical uncertainty. The lower pad shows the ratio of systematical variation to the nominal distribution. The black vertical error bars represent the MC statistical uncertainty. Same description applies to other envelope plots.
   }
\end{figure}

In the first case, we introduce the systematical uncertainty on the transverse momentum calibration of the hadronic tau. The corresponding nuisance parameter is denoted by $\theta(\pT(\tauhad))$. The uncertainty size is $\Delta \pT(\tauhad) = \pm 2\% \pT(\tauhad)$. It should be noted that the reconstruction of $\etmiss$ and all mass variables are also changed accordingly due to this uncertainty. 
Figure~\ref{fig:ex1_envelope_tes} shows the envelope plots to illustrate the effect of the systematical uncertainty on the BDT score distribution. An envelope plot shows systematical variations (there could high and low variations for each systematical item) on the background distribution as well as the nominal signal and background distributions.  But it seems not easy to draw conclusions on how different algorithms treat with the systematical uncertainty from these plots.

Both the statistics-only (stat.-only) fit, not including any systematical uncertainty, and the full fit with this systematical uncertainty are performed using a test statistics, based on the profile likelihood ratio defined in Ref.~\cite{asimov}, in the framework, HistoFactory~\cite{histofactory}. The expected signal significance is obtained from the distributions of the test statistics under the null (background-only) and alternative (signal-plus-background)
hypotheses using the asymptotic formulae in Ref.~\cite{asimov}. 
The numerical results for the expected signal significance and the fitted uncertainty of the nuisance parameter are summarized in Table~\ref{tab:ex1_sig}. We can see that the stat.-only significance is very close to each other. If the systematical uncertainty is included, QBDT0 has a similar significance as GradBDT, but QBDT1, the one considering the systematical uncertainty in training, gives the highest significance. 
We notice that $\theta(\tau \pT)$ is over-constrained with a similar degree for all BDT methods. But this is expected. As $\tau \pT$ itself is an input variable and other mass variables depends upon its reconstruction, the BDT score is thus sensitive to $\tau \pT$ as well if other systematical uncertainties are not considered. 
In Appendix~\ref{app:significance}, we try to estimate the significance with or without prior constraint on the nuisance parameter and present different representations of significance to consider the full uncertainty size. The conclusion does not change if the prior constraint on the uncertainty of $\tau \pT$ is removed.

\begin{figure}
   \includegraphics[width=0.32\textwidth]{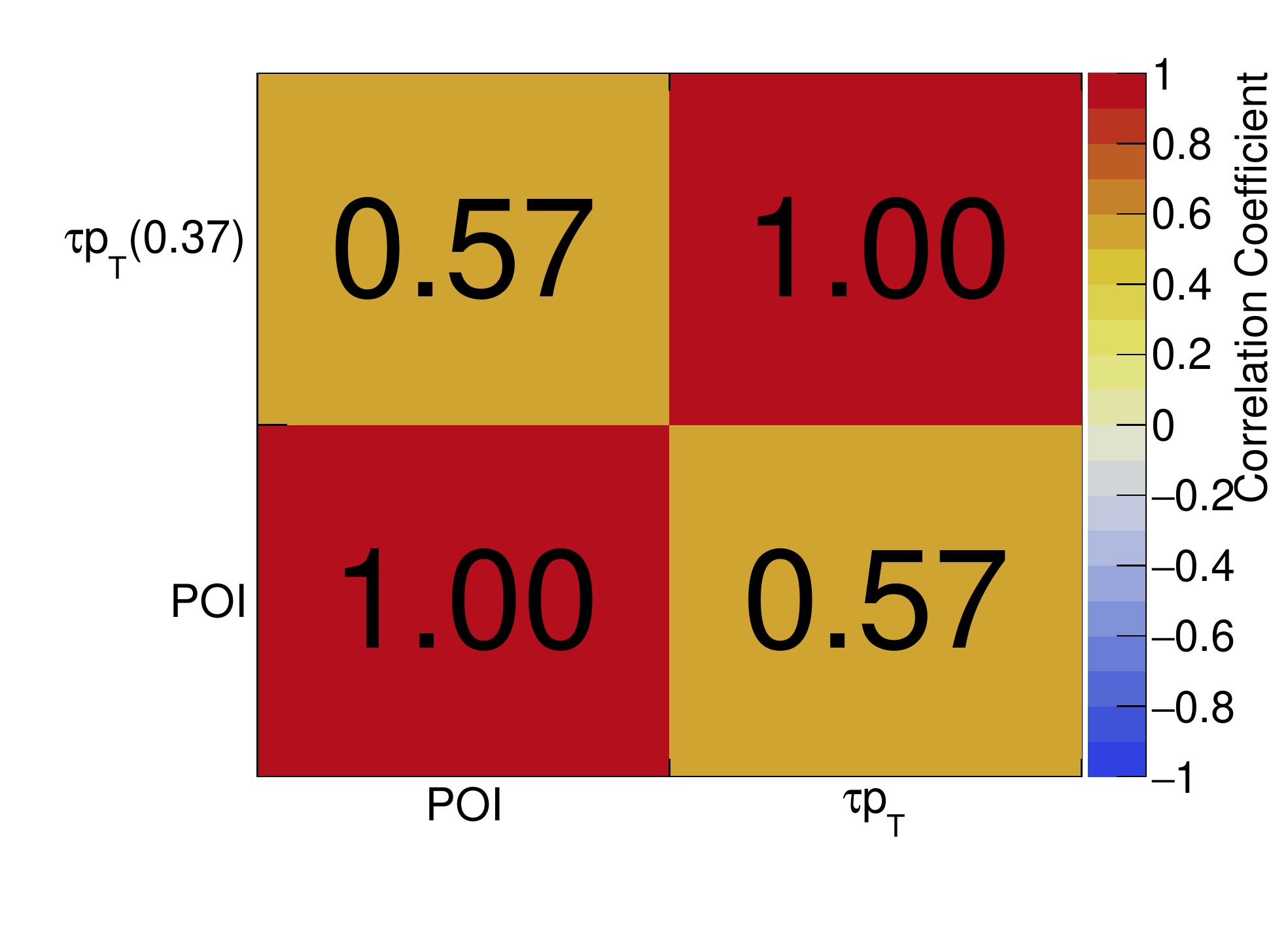}
   \includegraphics[width=0.32\textwidth]{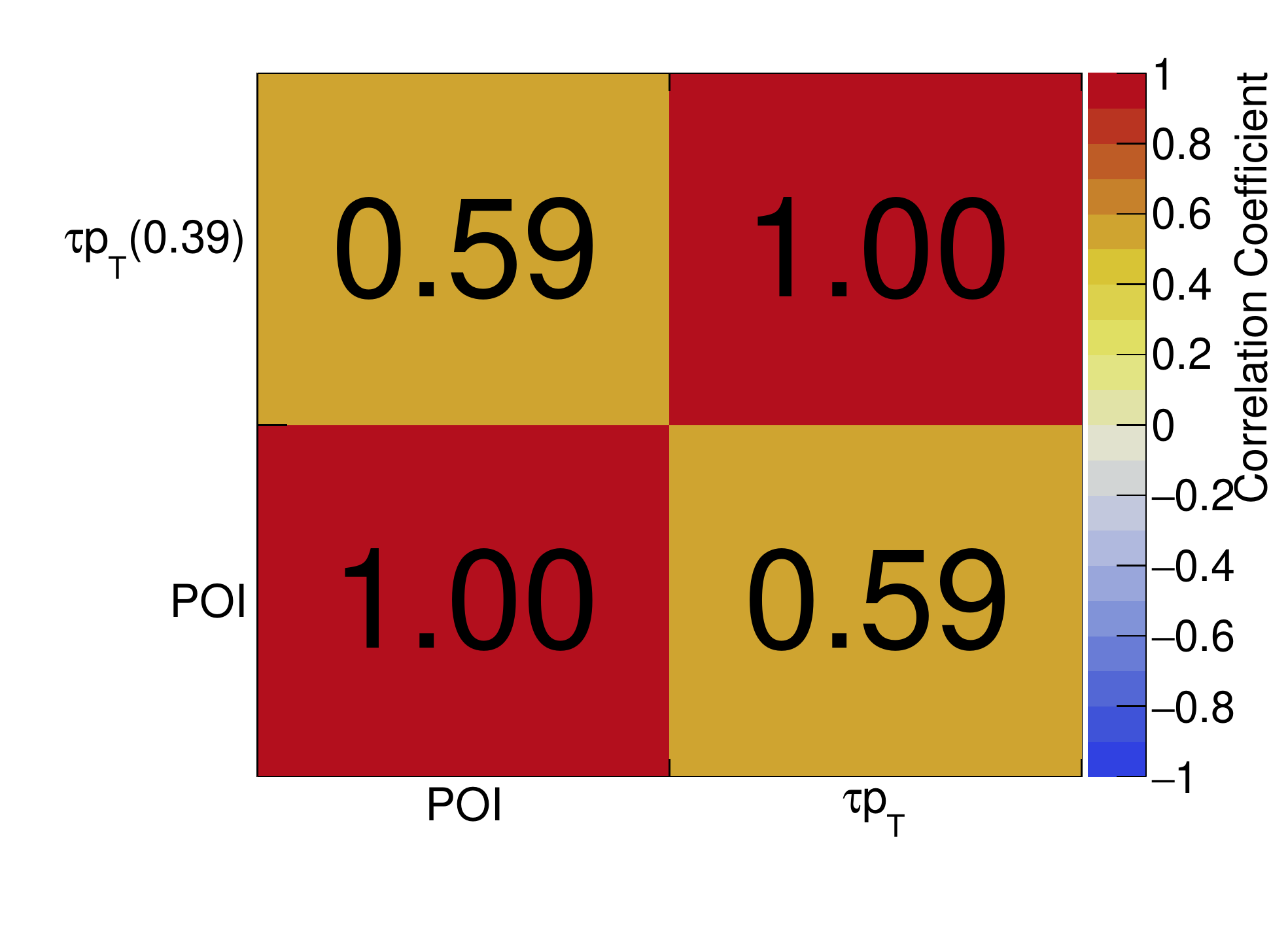}
   \includegraphics[width=0.32\textwidth]{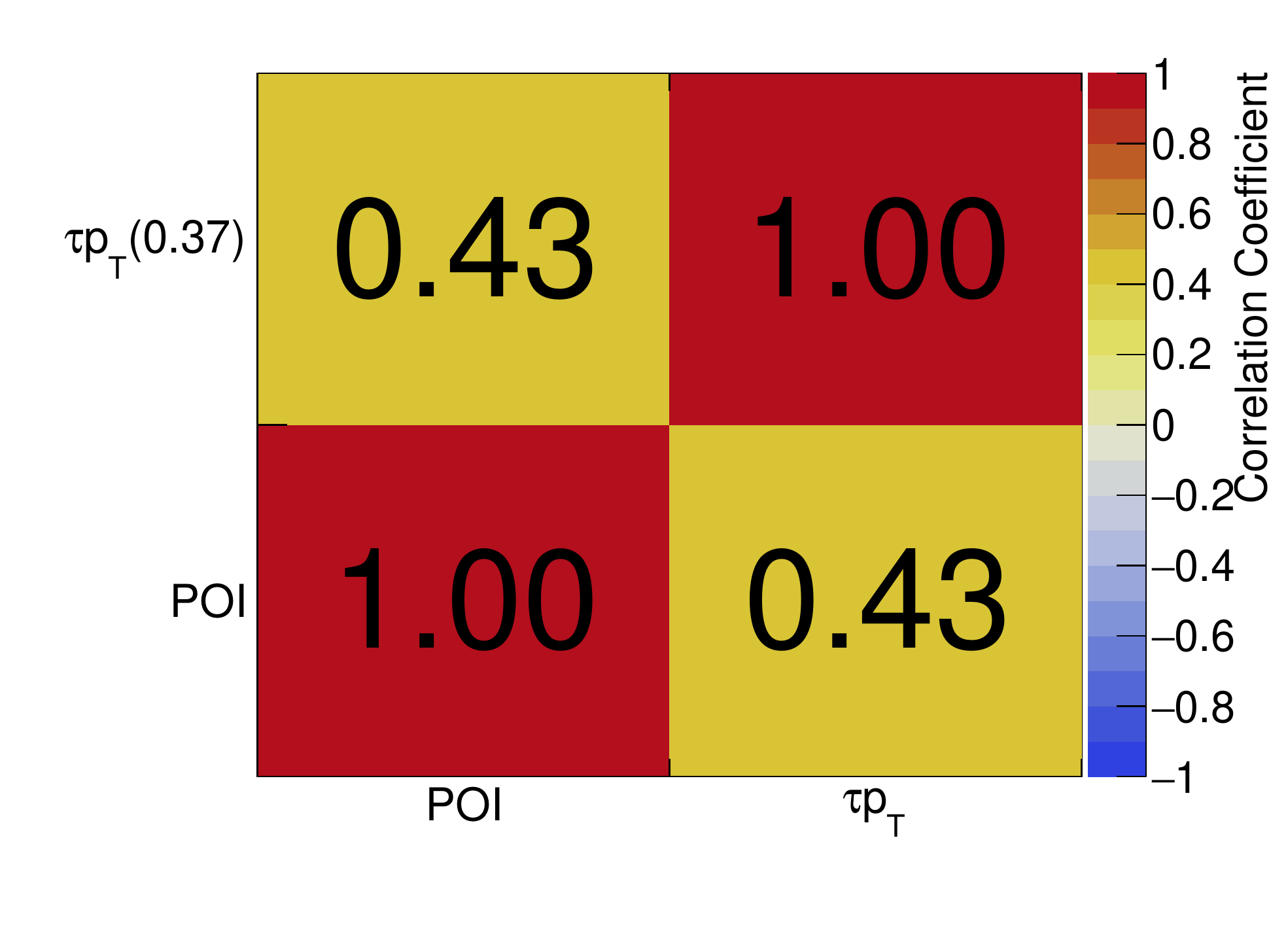}
   \caption{\label{fig:ex1_corr}
	Correlation matrix obtained from the fits. Left: GradBDT, middle: QBDT0 and right: QBDT1. 
   }
\end{figure}

One interesting and important feature happens to the correlation matrix shown in Fig.~\ref{fig:ex1_corr}. The correlation coefficient between the signal strength ( parameter of interest or POI ) and the systematical uncertainty source $\theta(\tau \pT)$ is $0.59$ for QBDT0, close to $0.57$ for GradBDT, but reduced to be $0.43$ in QBDT1. This is a good sign that QBDT1 is really learning about how to reduce the effect of the systematical uncertainty by decreasing the correlation between the signal strength and this systematical uncertainty source~\cite{constraint_xia}. 
As the correlation reduces, it is not surprising that the total significance is better.

\begin{table}
   \caption{\label{tab:ex1_sig} 
       Expected significance (unit: $\sigma$) expressed in number of standard deviation and post-fit uncertainty of the nuisance parameters. $\sigma_\theta$ is the post-fit uncertainty of the nuisance parameter $\theta$.
   }
   \begin{ruledtabular}
	\begin{tabular}{l | c c c}
        Significance ($\sigma$) & GradBDT & QBDT0 & QBDT1 \\
	   \hline
       Stat.-only fit  & 0.89 & 0.90 & 0.87 \\
       Full fit & 0.72 & 0.71 & 0.78 \\
	  \hline
	  $\sigma_\theta(\tau \pT)$ & 0.38 & 0.39 & 0.37 \\
	\end{tabular}
   \end{ruledtabular}
\end{table}

To compare QBDT and GradBDT quantitatively, we propose to measure the performance using the contribution to the uncertainty of the signal strength from the systematical uncertainty sources. Let $\Delta\mu_0$ denote the uncertainty of the signal strength from the stat-only fit and $\Delta\mu_1$ denote that from the full fit. The contribution from the systematical uncertainties can be represented by $\Delta\mu$ defined as $\Delta\mu\equiv\sqrt{\Delta\mu_1^2 -\Delta\mu_0^2}$
and the numerical results are summarized in Table~\ref{tab:ex3_dmu} for this case and the following two cases. We can see that the $\Delta\mu$ for QBDT1 is 71~\% of that for GradBDT.  

\begin{table}
   \caption{\label{tab:ex3_dmu} 
    Contributions to the uncertainty of the signal strength from the systematical uncertainties. The rightmost column is the ratio of the contribution in QBDTX and that in GradBDT.
   }
   \begin{ruledtabular}
     \begin{tabular}{l  l l l l}
	  $\Delta\mu$ & $\Nsysts$ &GradBDT & QBDTX & $\frac{\Delta\mu(\text{QBDTX})}{\Delta\mu(\text{GradBDT})}$ \\
	  \hline
	  Case I & 1 & 0.79 & 0.56 &  71\% \\ 
	  Case II& 3 & 0.82 & 0.71 & 85\% \\ 
	  Case III& 7 & 0.99 & 0.49 & 50\% \\ 
     \end{tabular}
   \end{ruledtabular}
\end{table}

\subsection{Case II: three systematical uncertainty sources}~\label{sec:ex3}

\begin{figure}
   \includegraphics[width=0.32\textwidth]{Cs_SR_tes_QBDTh2atataBDTGv8sevensystsSsp.pdf}
   \includegraphics[width=0.32\textwidth]{Cs_SR_tes_QBDTh2atataQBDT0v8sevensystsSsp.pdf}
   \includegraphics[width=0.32\textwidth]{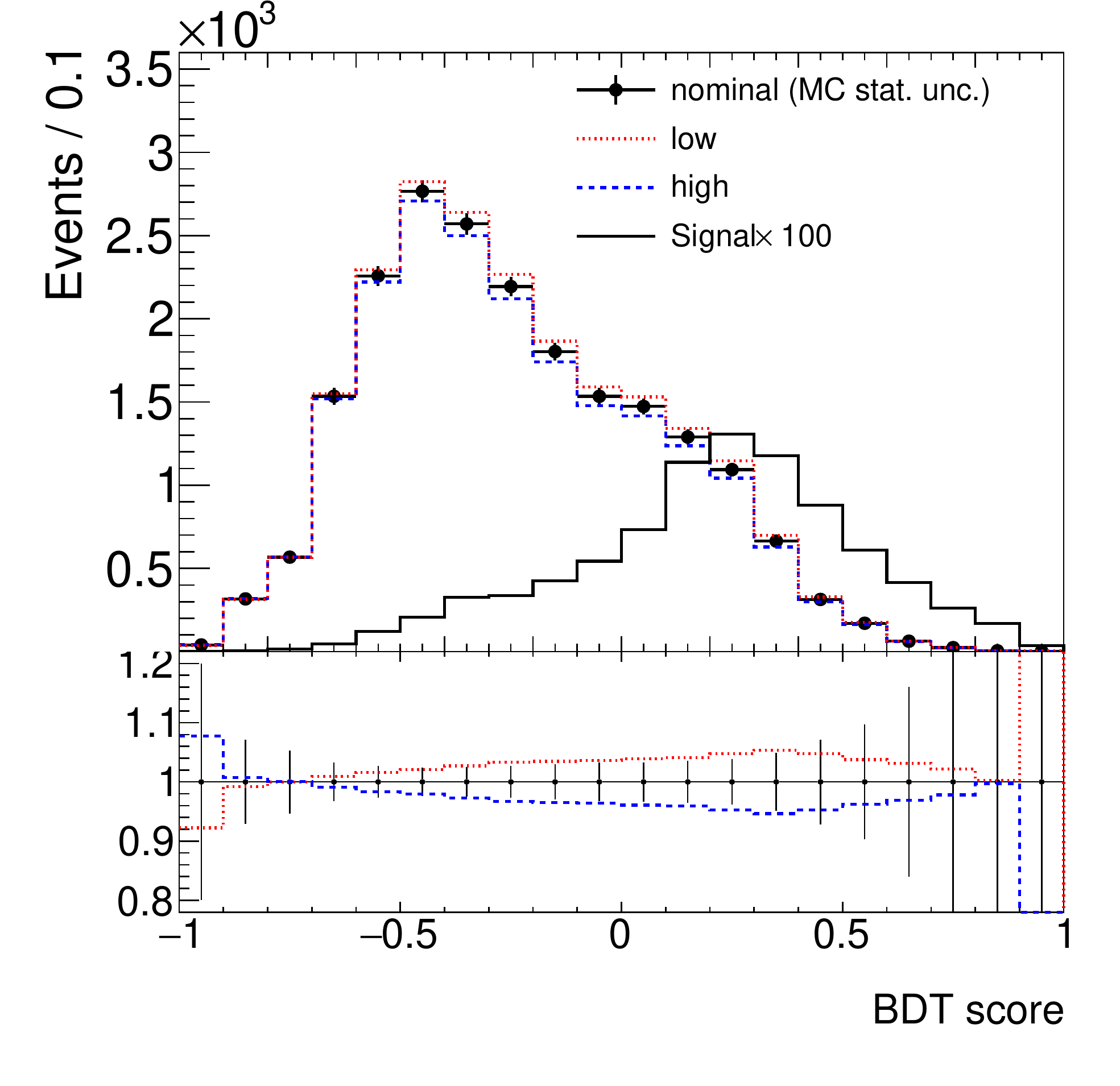}
   \caption{\label{fig:ex3_envelope_tes}
       (color online) Envelope plot for the systematical uncertainty of tau energy scale. Left: GradBDT, middle: QBDT0 and right: QBDT3. 
   }
\end{figure}

\begin{figure}
   \includegraphics[width=0.32\textwidth]{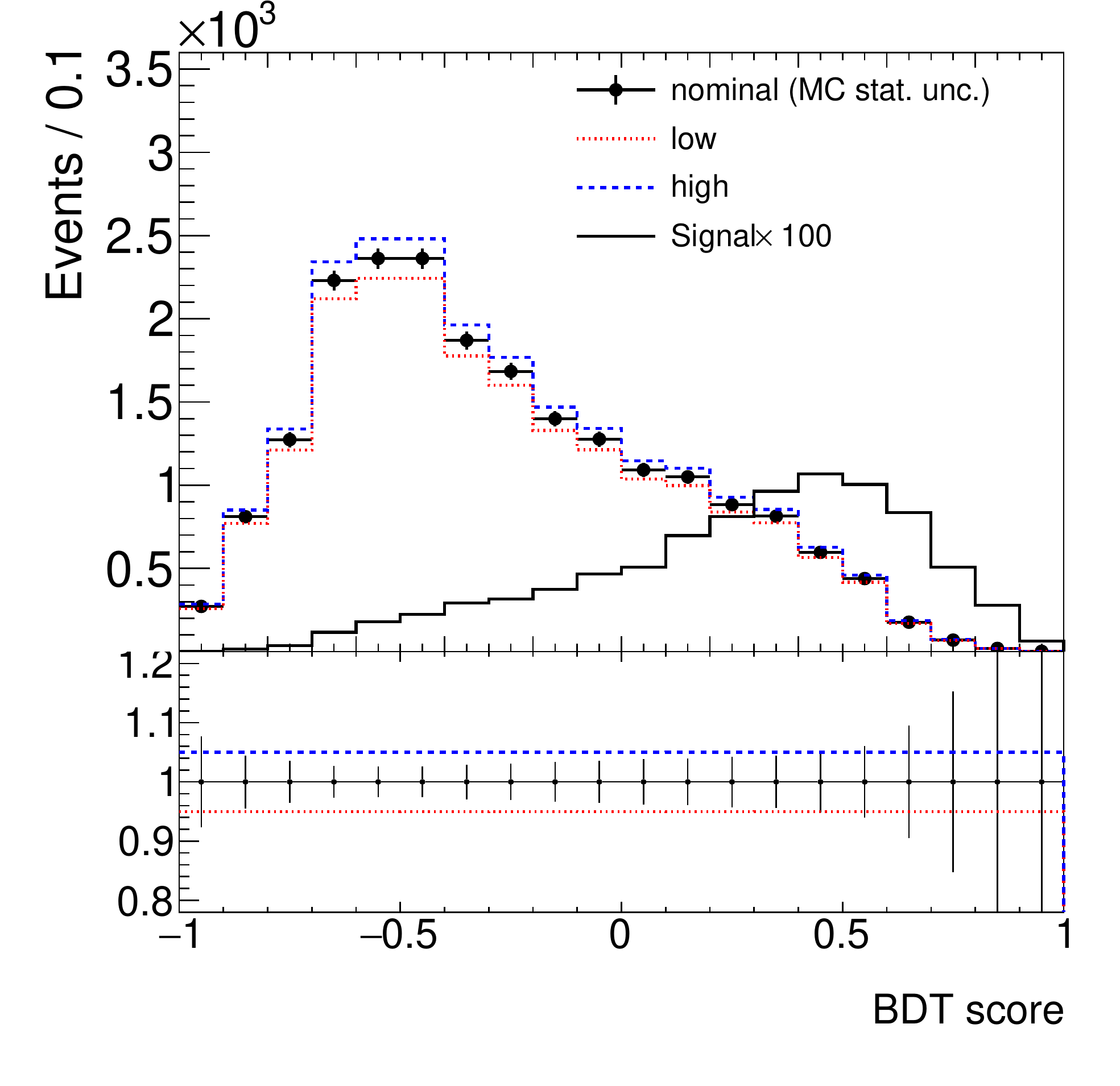}
   \includegraphics[width=0.32\textwidth]{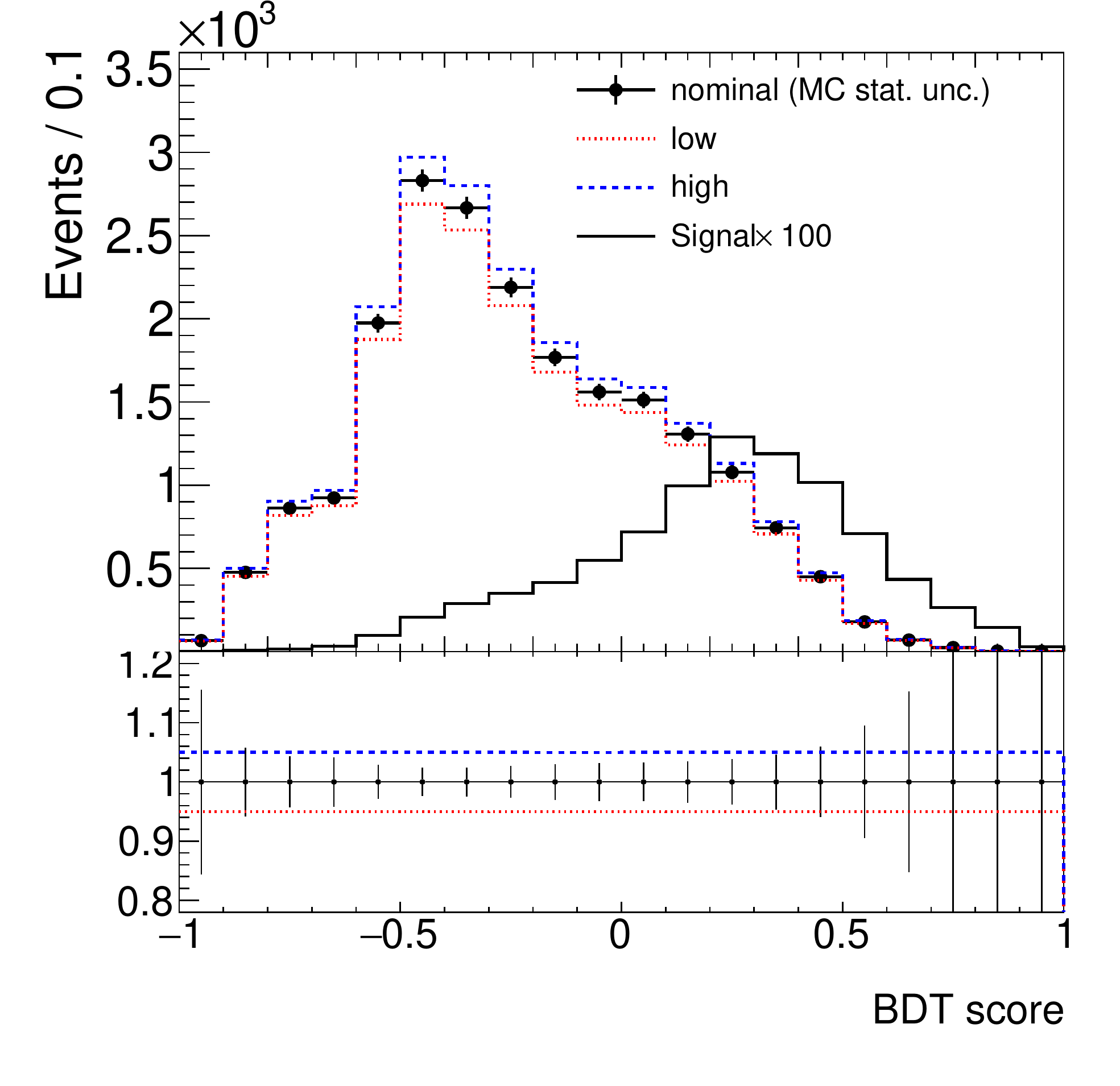}
   \includegraphics[width=0.32\textwidth]{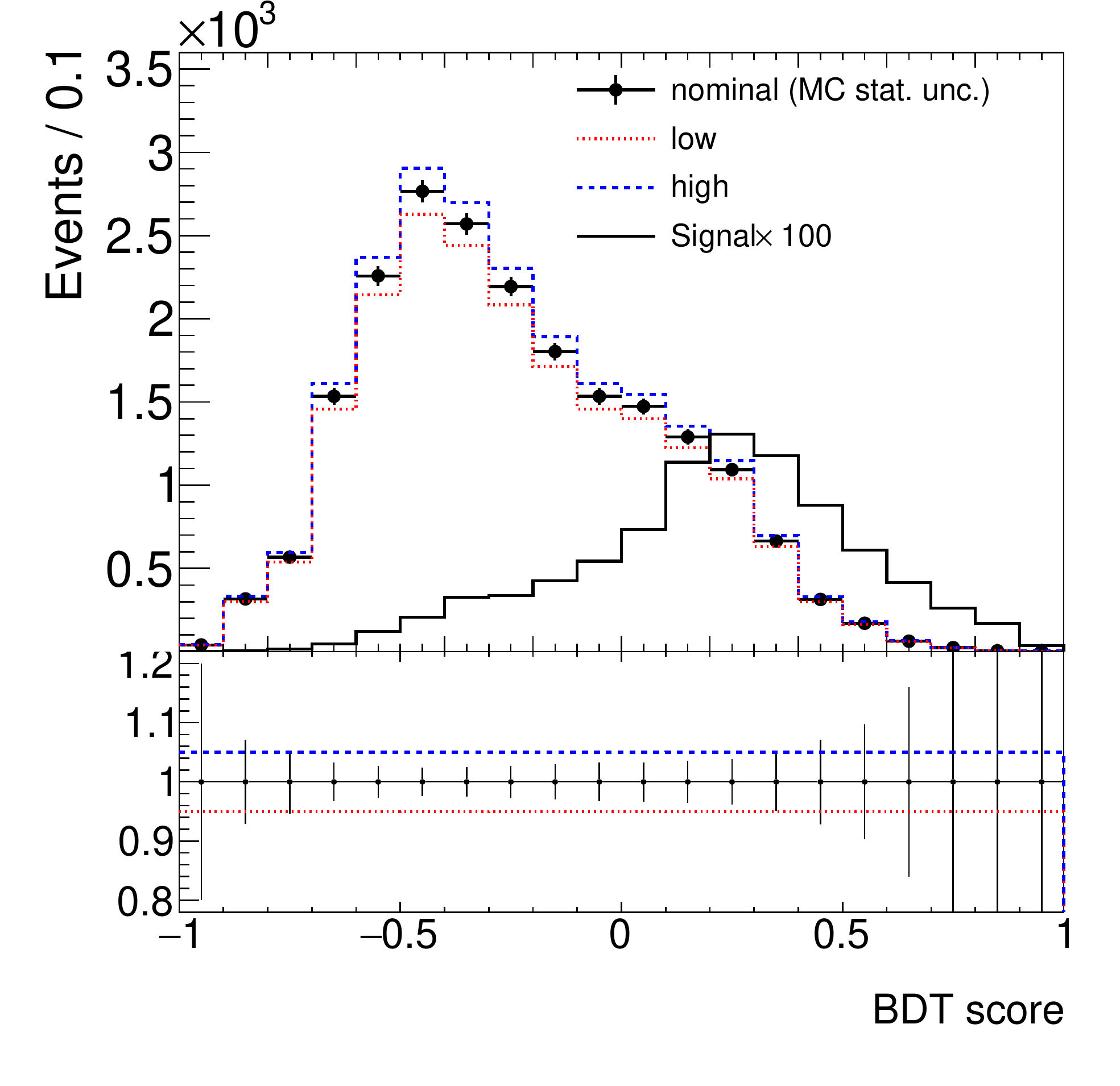}
   \caption{\label{fig:ex3_envelope_tauid}
   (color online) Envelope plot for the systematical uncertainty of tau ID efficiency. Left: GradBDT, middle: QBDT0, and right: QBDT3. }
\end{figure}

\begin{figure}
   \includegraphics[width=0.32\textwidth]{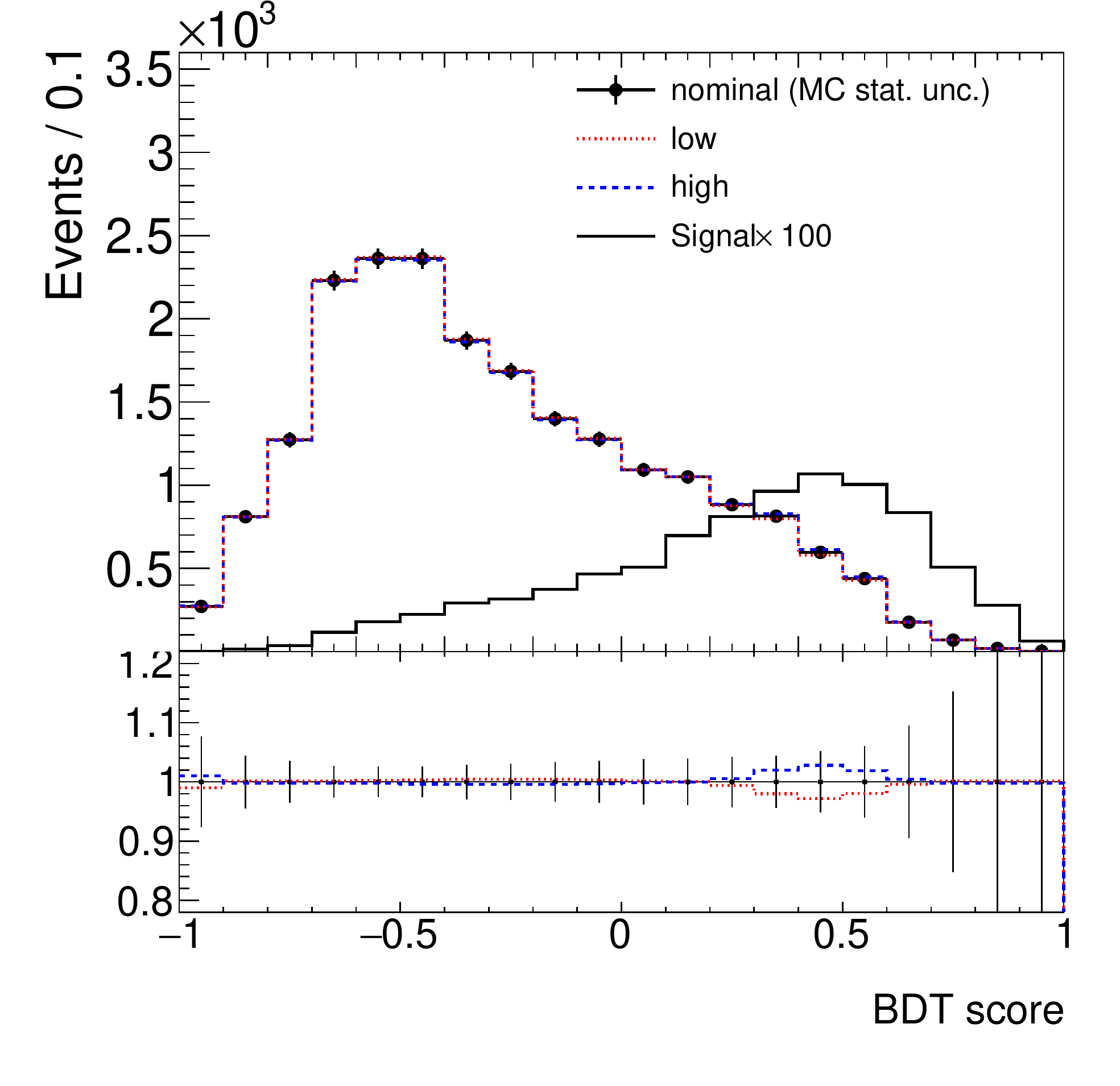}
   \includegraphics[width=0.32\textwidth]{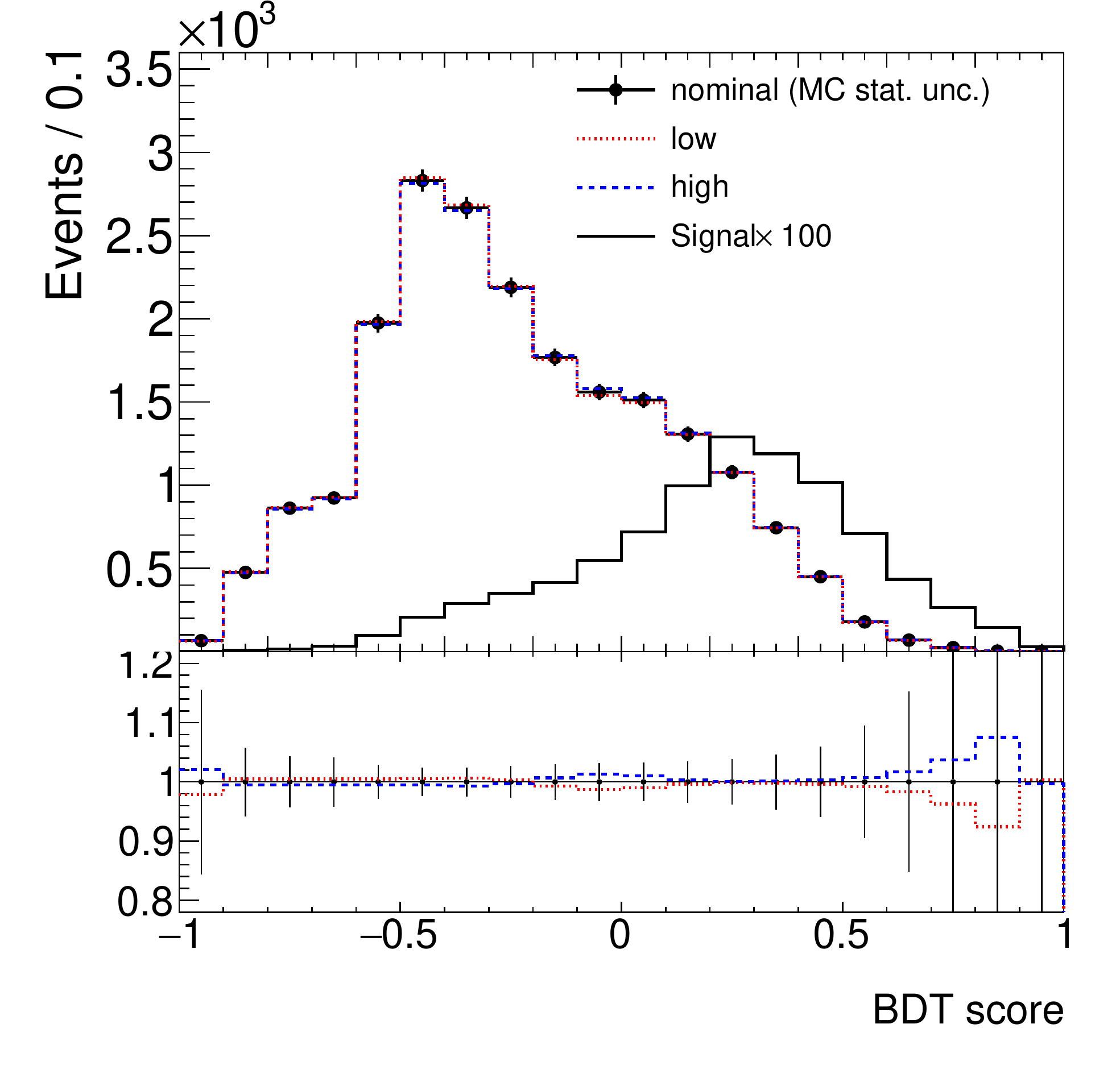}
   \includegraphics[width=0.32\textwidth]{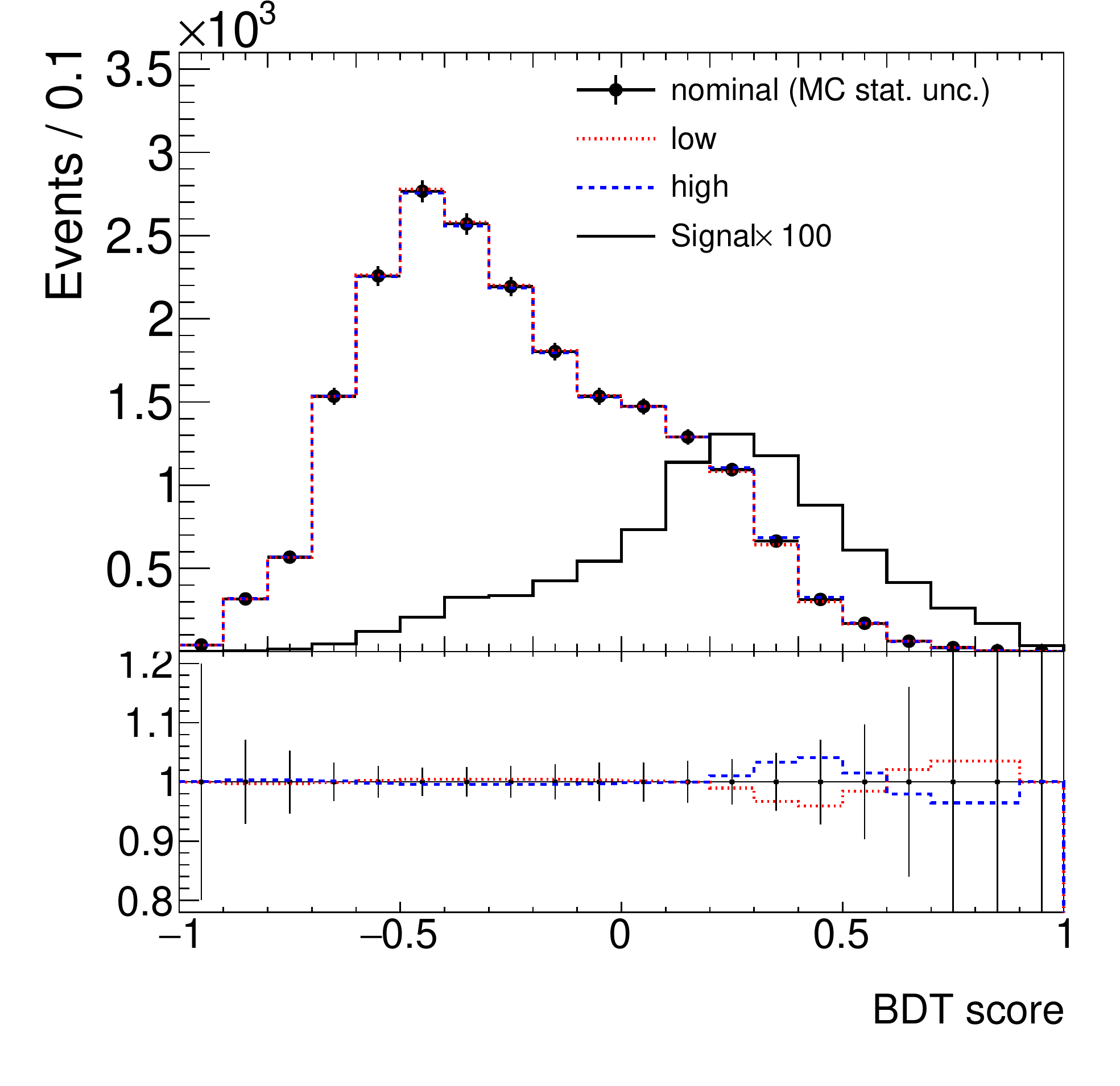}
   \caption{\label{fig:ex3_envelope_met}
       (color online) Envelope plot for the systematical uncertainty of missing transverse energy resoltuion. Left: GradBDT, middle: QBDT0, and right: QBDT3. 
   }
\end{figure}

In this case, two more systematical uncertainty sources are introduced. One is the $\tauhad$ identification efficiency uncertainty. It is $5\%$ independent upon the transverse momentum. The other is $\etmiss$ resolution uncertainty of 10\%, which is applied to both its magnitude and direction. The envelope plots for these systematical uncertainties are shown in Fig.~\ref{fig:ex3_envelope_tes},~\ref{fig:ex3_envelope_tauid} and ~\ref{fig:ex3_envelope_met}, respectively. Note that for the tau $\pT$ calibration uncertainty, the envelope plots from GradBDT and BDT0 in Fig.~\ref{fig:ex3_envelope_tes} are the same as in Case I (namely, those in Fig.~\ref{fig:ex1_envelope_tes}). 

\begin{figure}
   \includegraphics[width=0.32\textwidth]{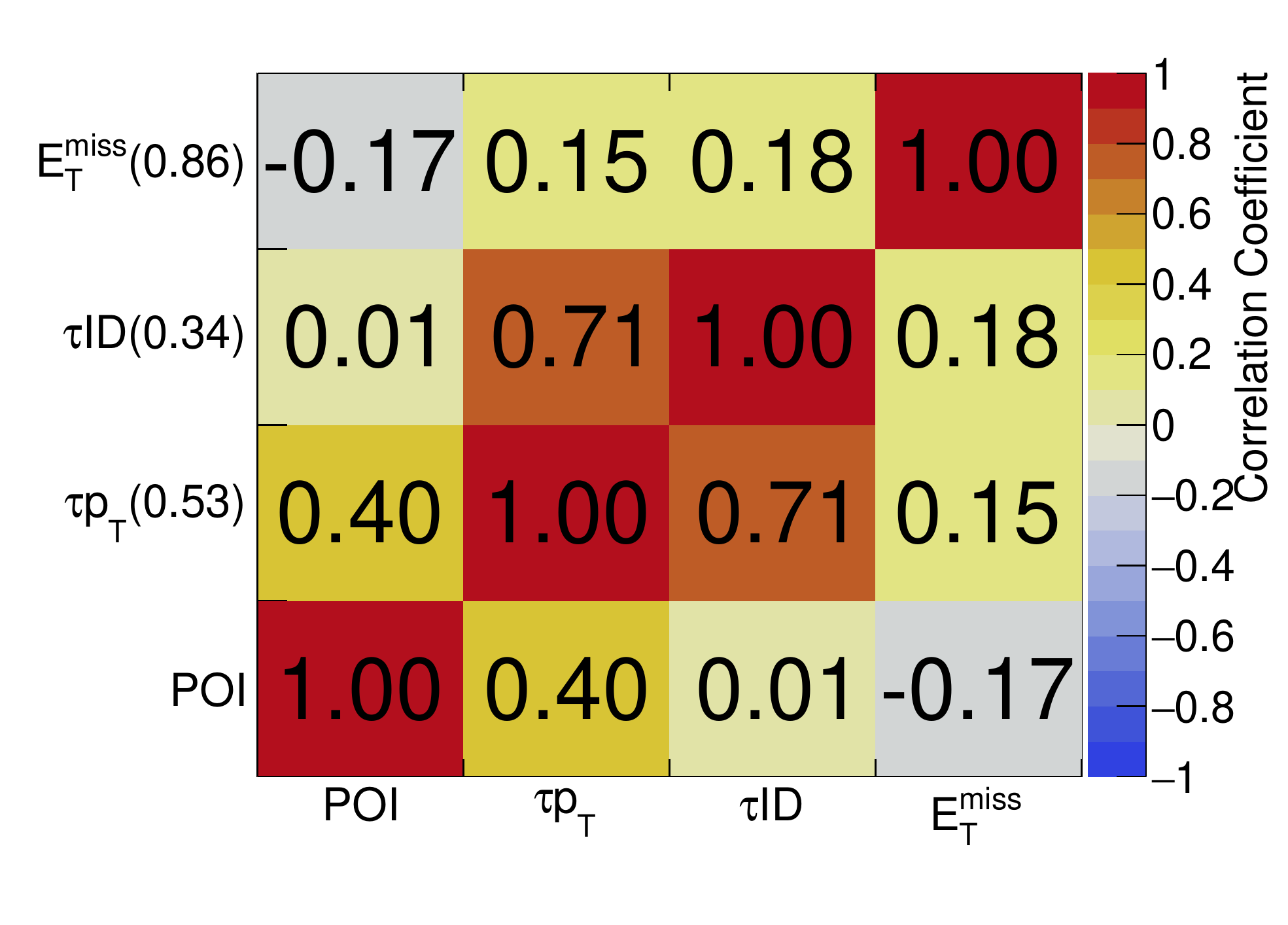}
   \includegraphics[width=0.32\textwidth]{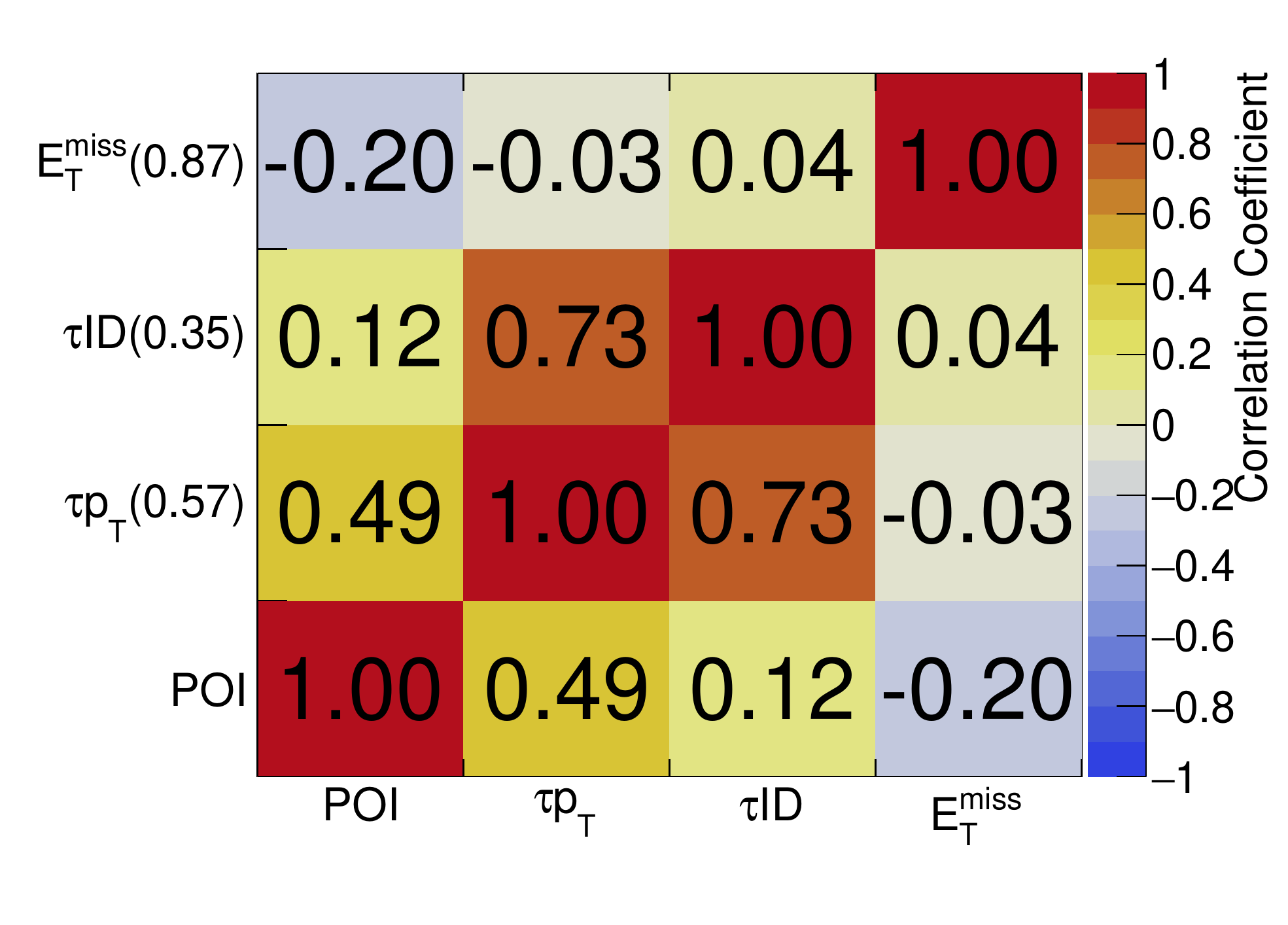}
   \includegraphics[width=0.32\textwidth]{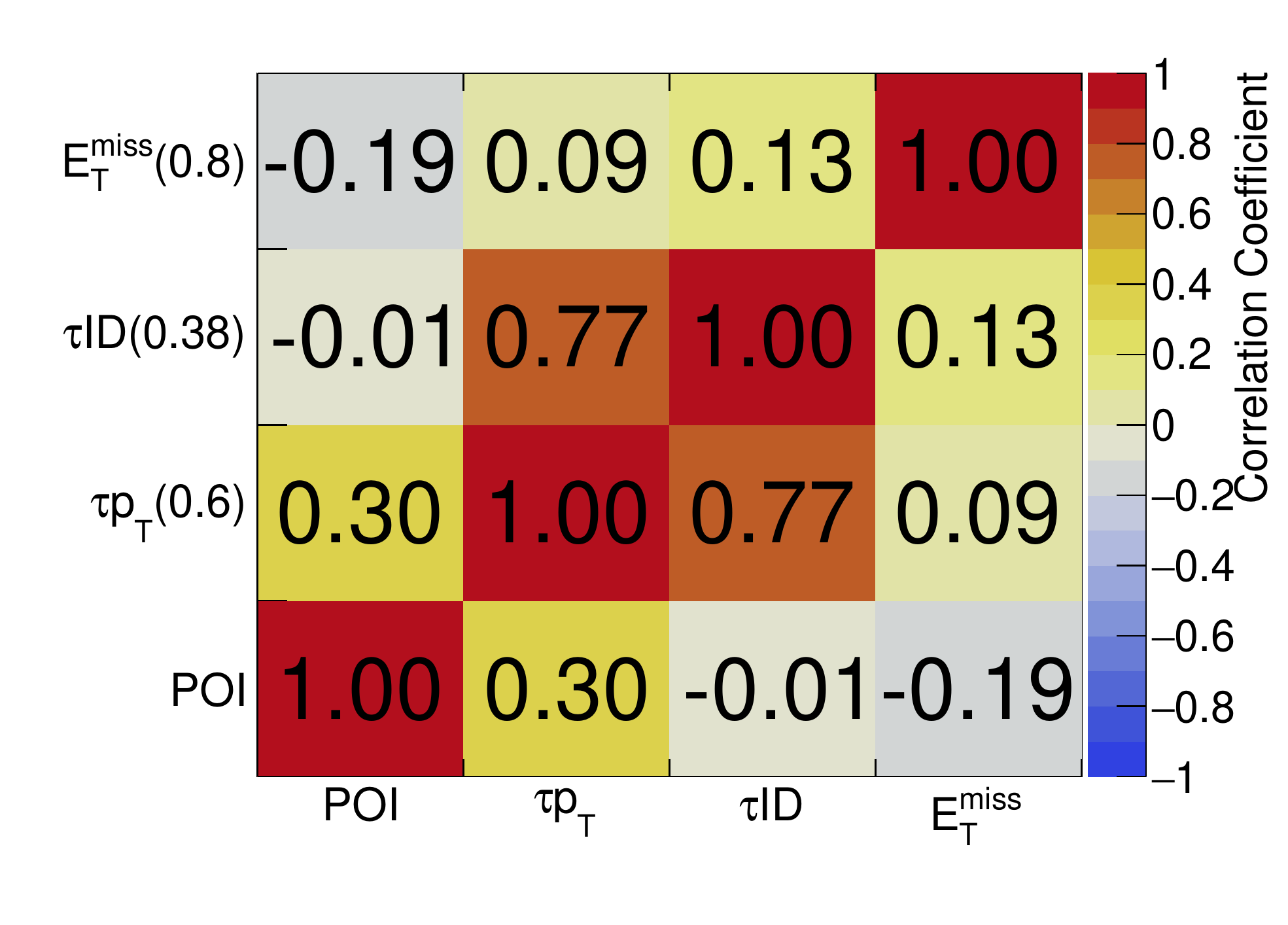}
   \caption{\label{fig:ex3_corr}
	Correlation matrix obtained from the fits. Left: GradBDT, middle: QBDT0, and right: QBDT3. 
   }
\end{figure}

The fit results are summarized in Table~\ref{tab:ex3_sig}. We can draw the same conclusion as in Case~I that GradBDT and QBDT0 have similar performance, and QBDT3 is the best. We do not see all nuisance parameters are constrained in any specific BDT setup.
The correlation matrices are compared in Fig.~\ref{fig:ex3_corr}. We can also see that the correlation between the signal strength and other nuisance parameters is reduced in QBDT3. The signal strength uncertainty $\Delta\mu$ from the systematical uncertainties in QBDT is 85\% of that in GradBDT as shown in Table~\ref{tab:ex3_dmu}. This fraction is worse than that in Case~I. 
One possible reason is that the tau ID efficiency uncertainty only affects the background normalization and it seems difficult to improve the significance by optimizing the shape of the score distribution.

\begin{table}
   \caption{\label{tab:ex3_sig} 
       Expected significance (unit: $\sigma$) expressed in number of standard deviation and post-fit uncertainty of the nuisance parameters. $\sigma_\theta$ is the post-fit uncertainty of the nuisance parameter $\theta$.
   }
   \begin{ruledtabular}
     \begin{tabular}{l | c c c}
         Significance ($\sigma$)  & GradBDT & QBDT0 & QBDT3 \\
        \hline
	  Stat.-only fit & 0.89 & 0.90 & 0.89 \\
	  Full fit & 0.71 & 0.69 & 0.75 \\
	  \hline
	  $\sigma_\theta(\tau \pT)$ & 0.53 & 0.57 & 0.61\\
	  $\sigma_\theta(\tau ID)$ & 0.34 & 0.35 & 0.38 \\
	  $\sigma_\theta(E_T^{\text{miss}})$ & 0.87 & 0.87 & 0.80 \\
     \end{tabular}
   \end{ruledtabular}
\end{table}

   \subsection{Case III: seven systematical uncertainty sources}~\label{sec:ex7}
   Here in this case 7 systematical uncertainty sources are considered. The definitions have been already summarized in Table~\ref{tab:ex1_systs}. Except those introduced above, we have the lepton $\pT$ calibration uncertainty, lepton identification efficiency uncertainty, photon $\pT$ calibration uncertainty and photon identification efficiency uncertainty. 
   
\begin{figure}
	\includegraphics[width=0.35\textwidth]{Cs_SR_tes_QBDTh2atataBDTGv8sevensystsSsp.pdf}
	\includegraphics[width=0.35\textwidth]{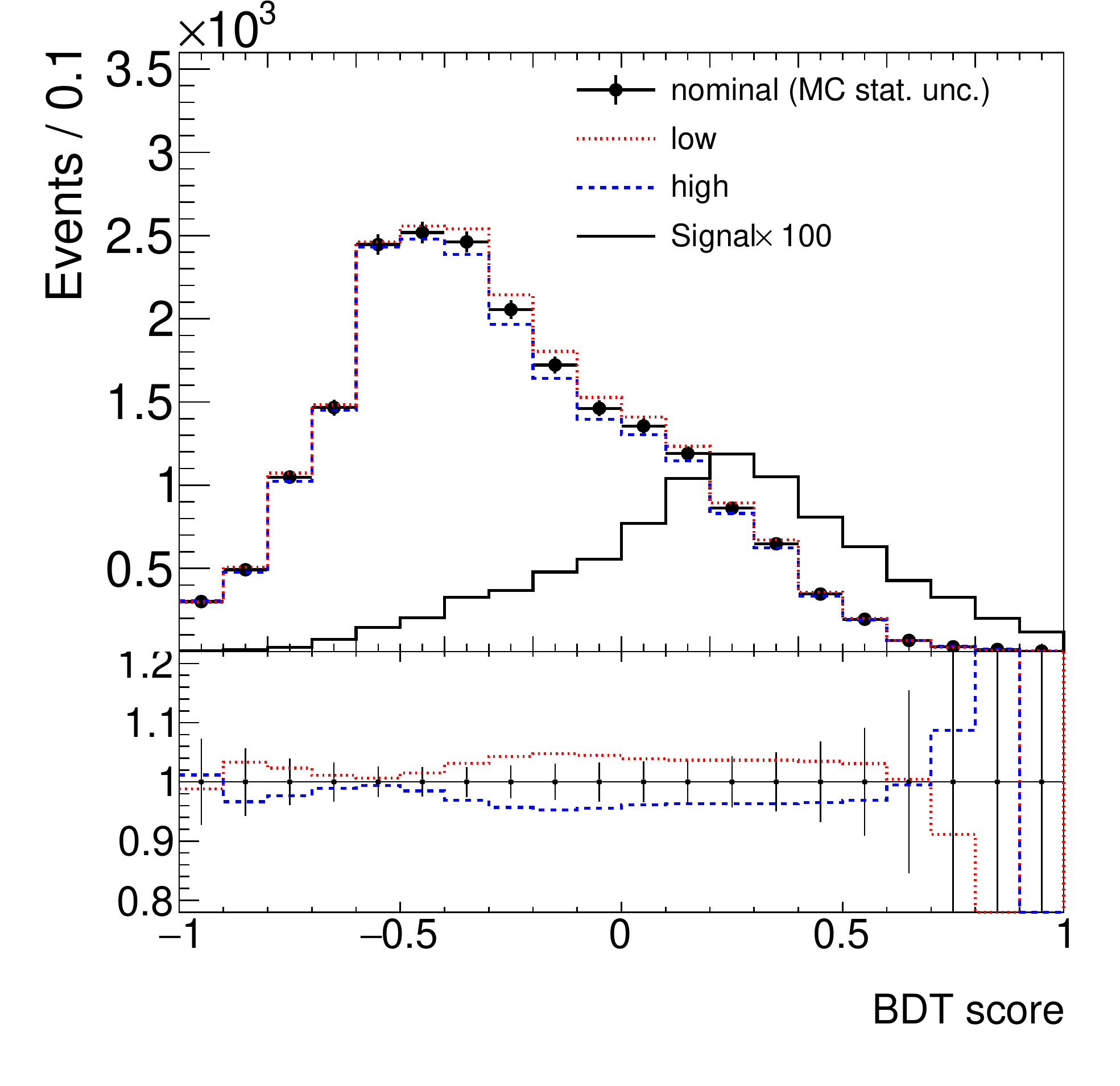}\\
	\includegraphics[width=0.35\textwidth]{Cs_SR_tauid_QBDTh2atataBDTGv8sevensystsSsp.pdf}
	\includegraphics[width=0.35\textwidth]{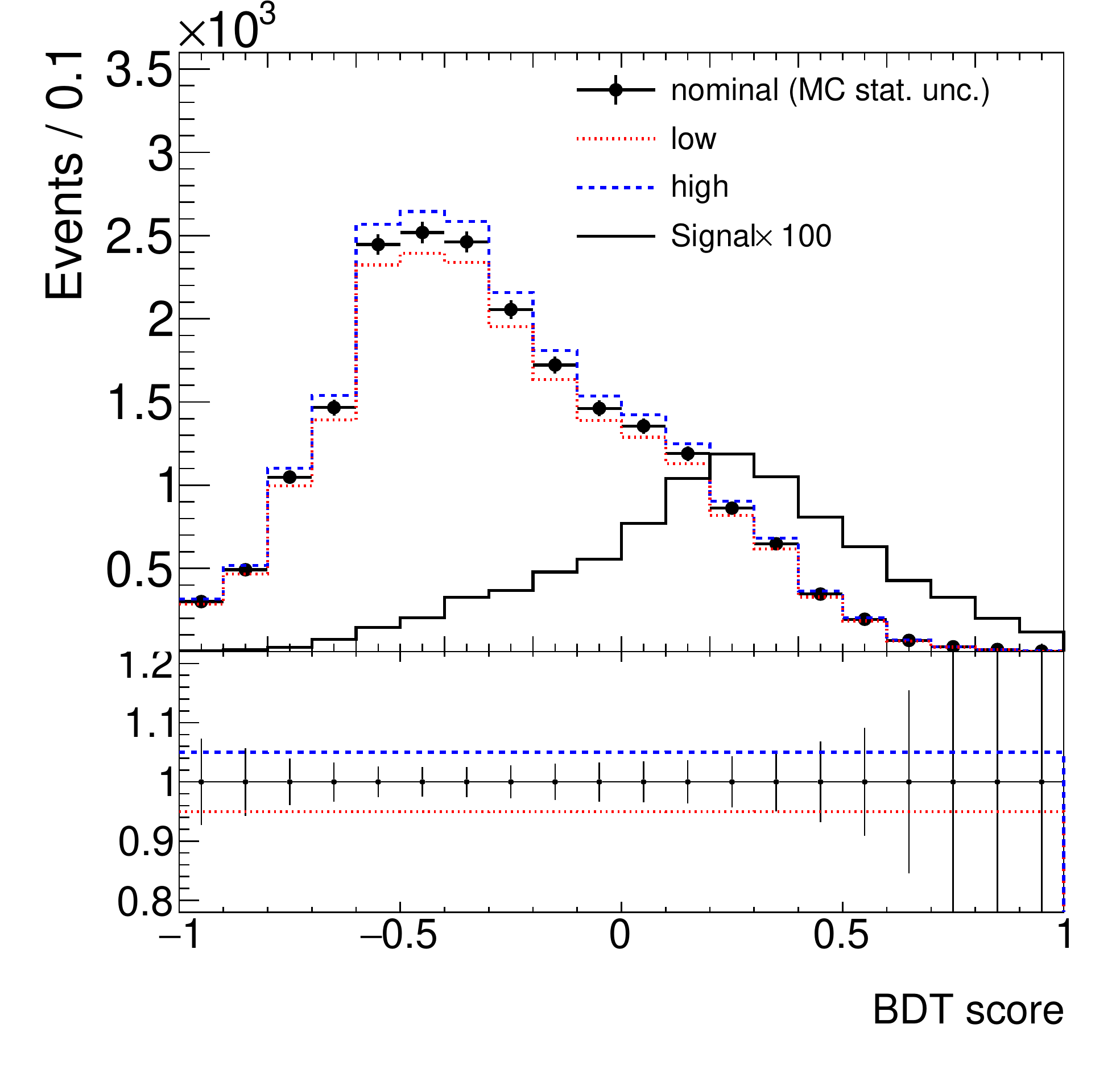}\\
	\includegraphics[width=0.35\textwidth]{Cs_SR_met_QBDTh2atataBDTGv8sevensystsSsp.pdf}
	\includegraphics[width=0.35\textwidth]{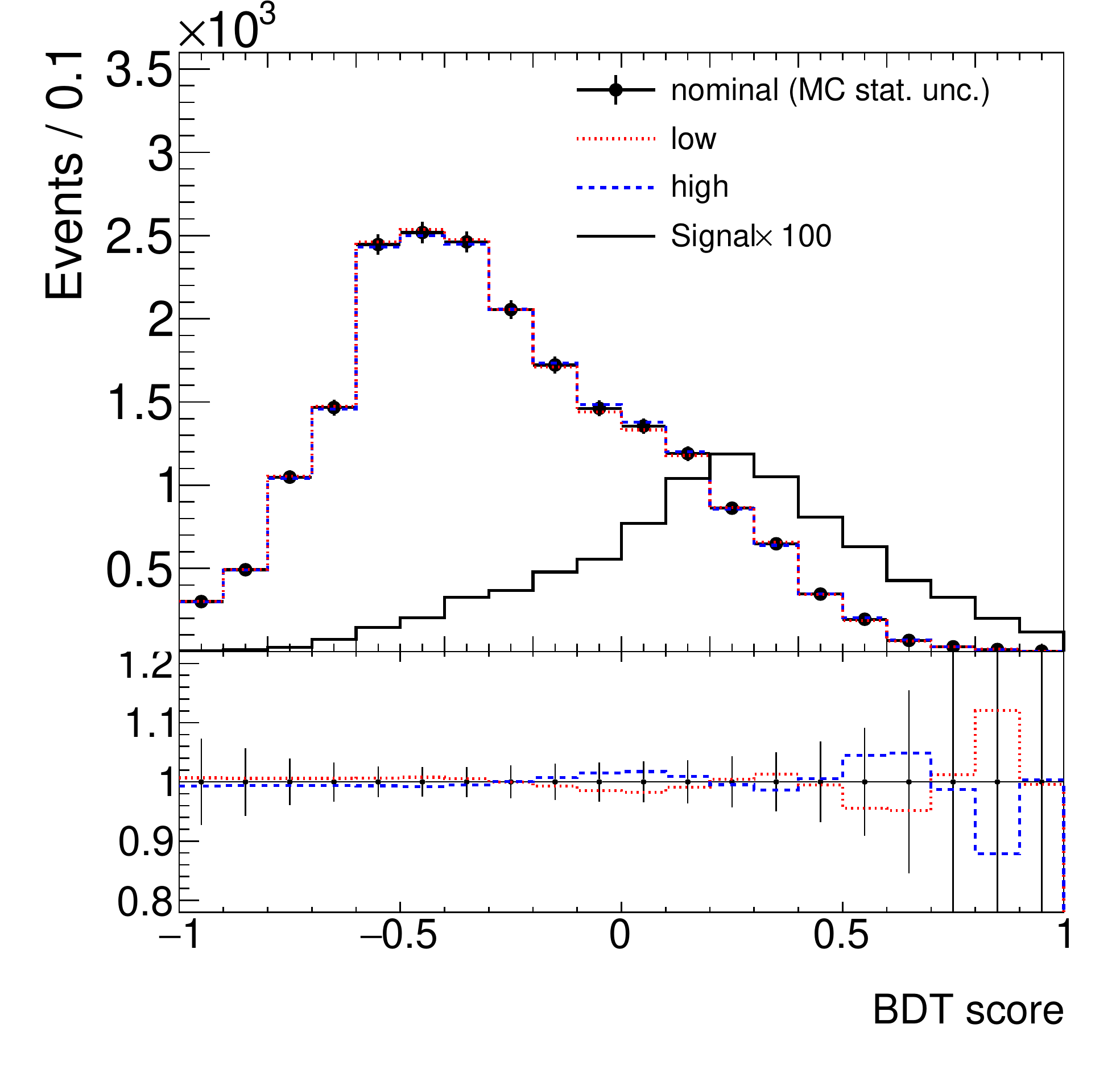}
	\caption{\label{fig:ex7_envelope_part1}
        (color online) Envelope plot for the systematical uncertainties. Left column: GradBDT, right column: QBDT7. From top to bottom, the systematical uncertainty source is tau $\pT$ calibration, tau ID efficiency and $\etmiss$ resolution. 
	}
\end{figure}

\begin{figure}
	\includegraphics[width=0.35\textwidth]{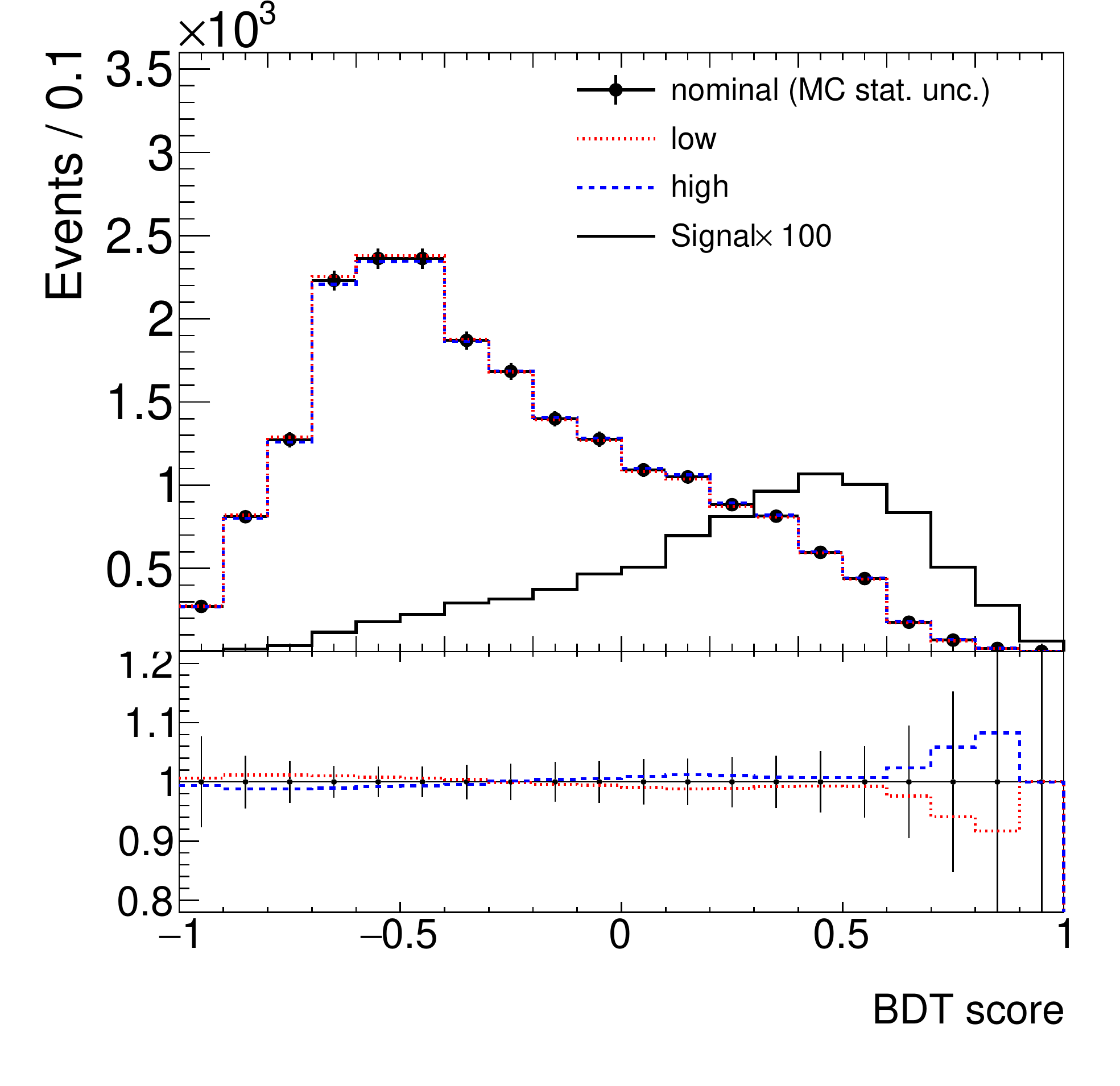}
	\includegraphics[width=0.35\textwidth]{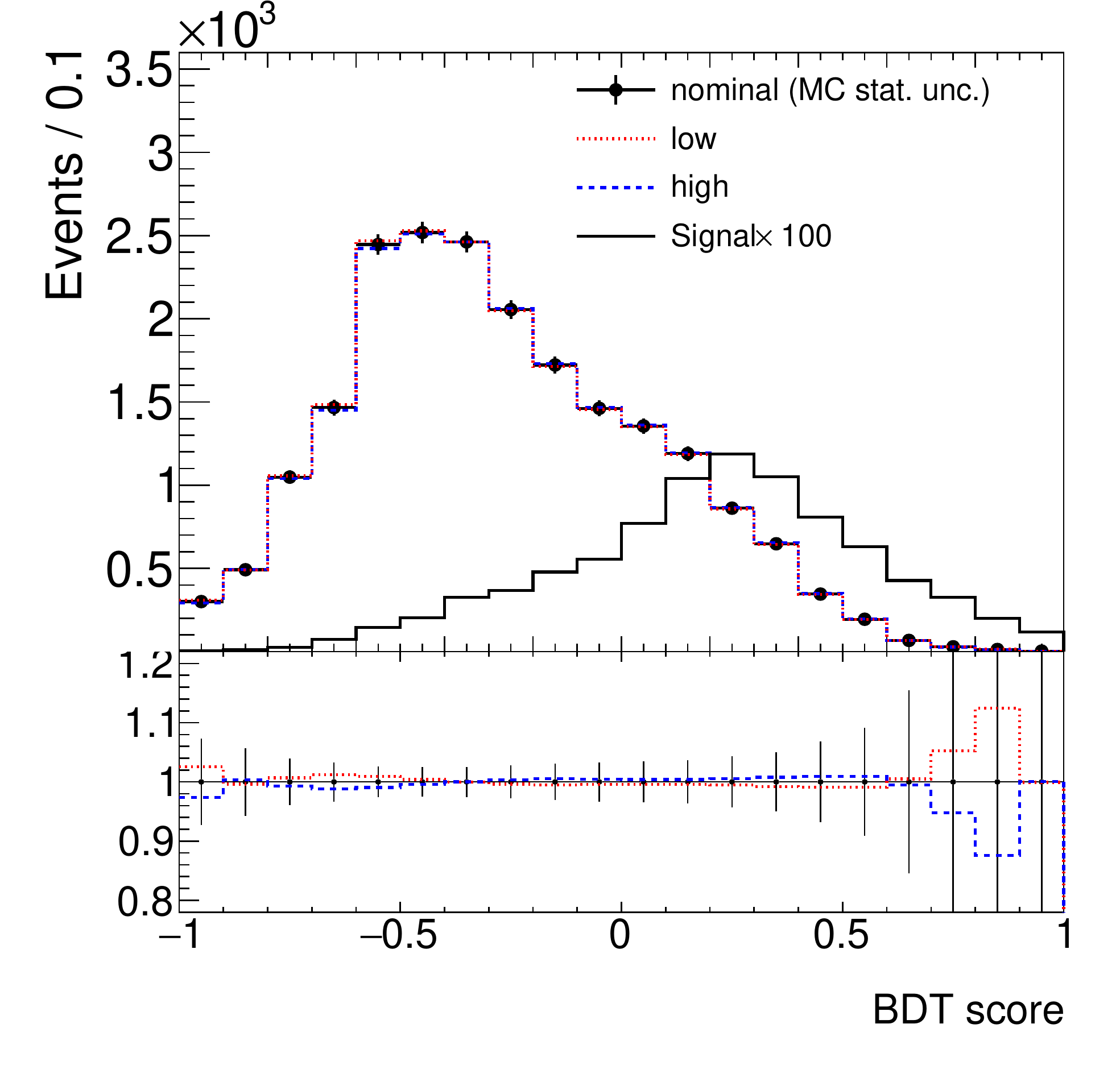}\\
	\includegraphics[width=0.35\textwidth]{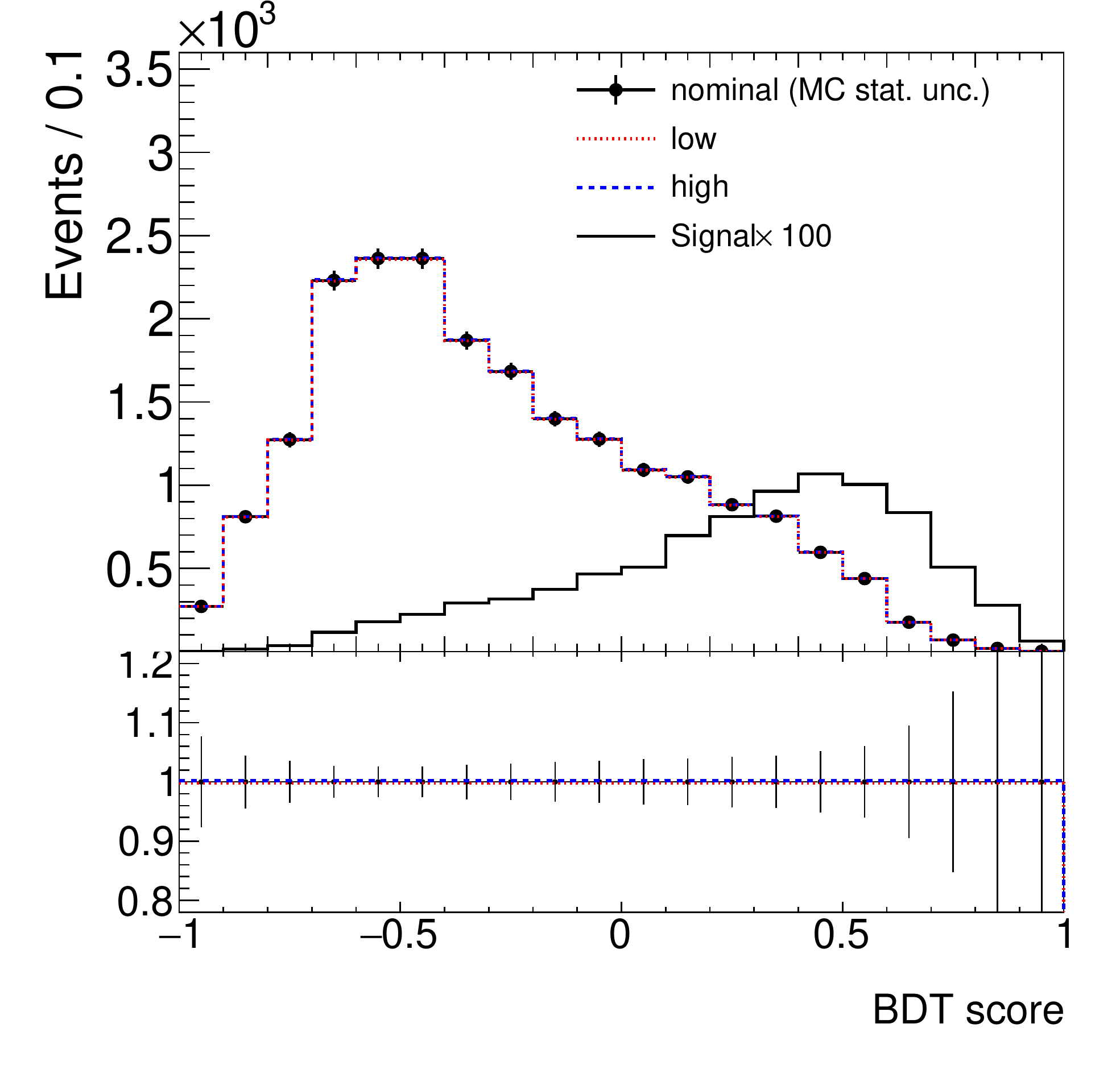}
	\includegraphics[width=0.35\textwidth]{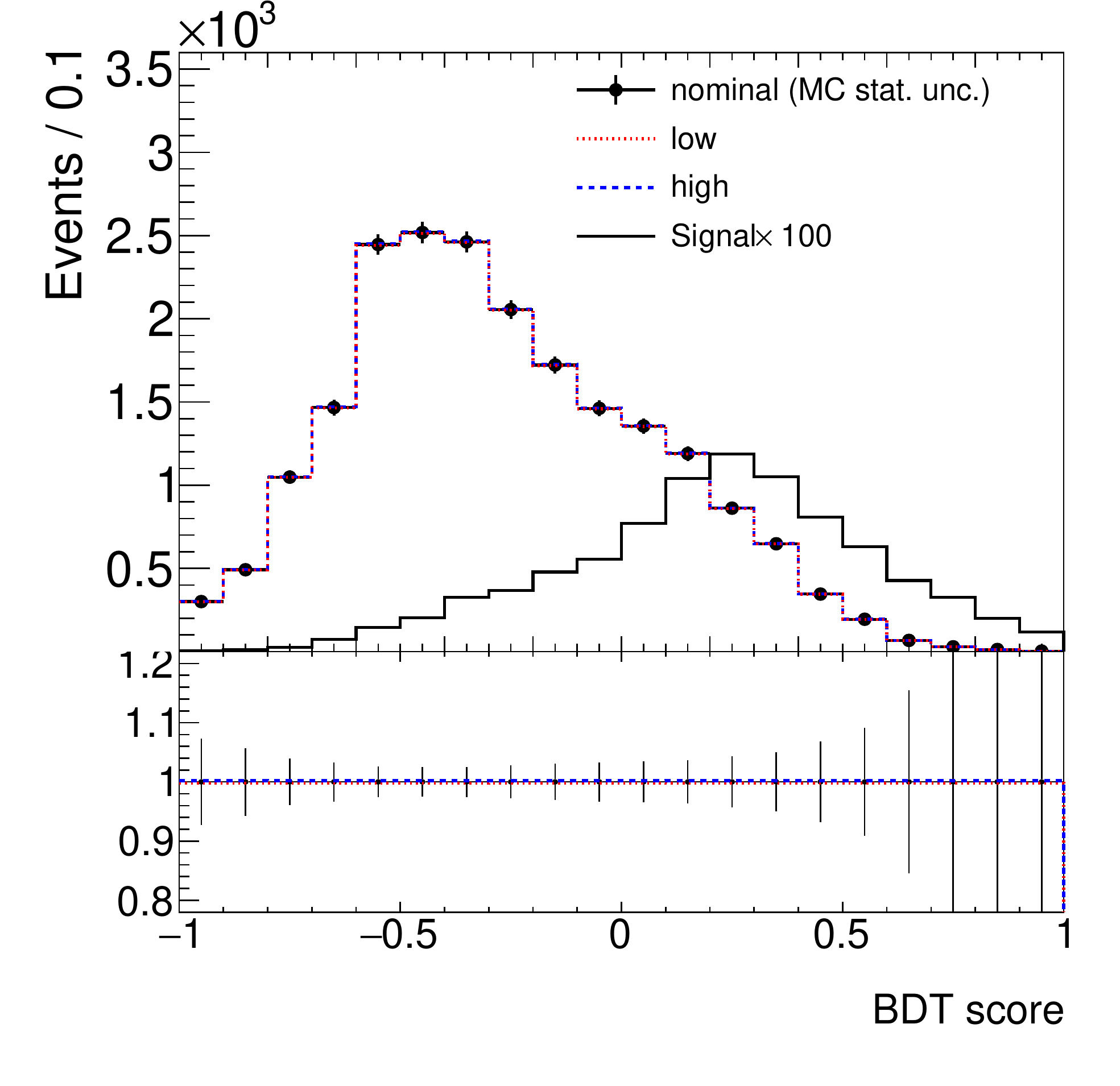}\\
	\includegraphics[width=0.35\textwidth]{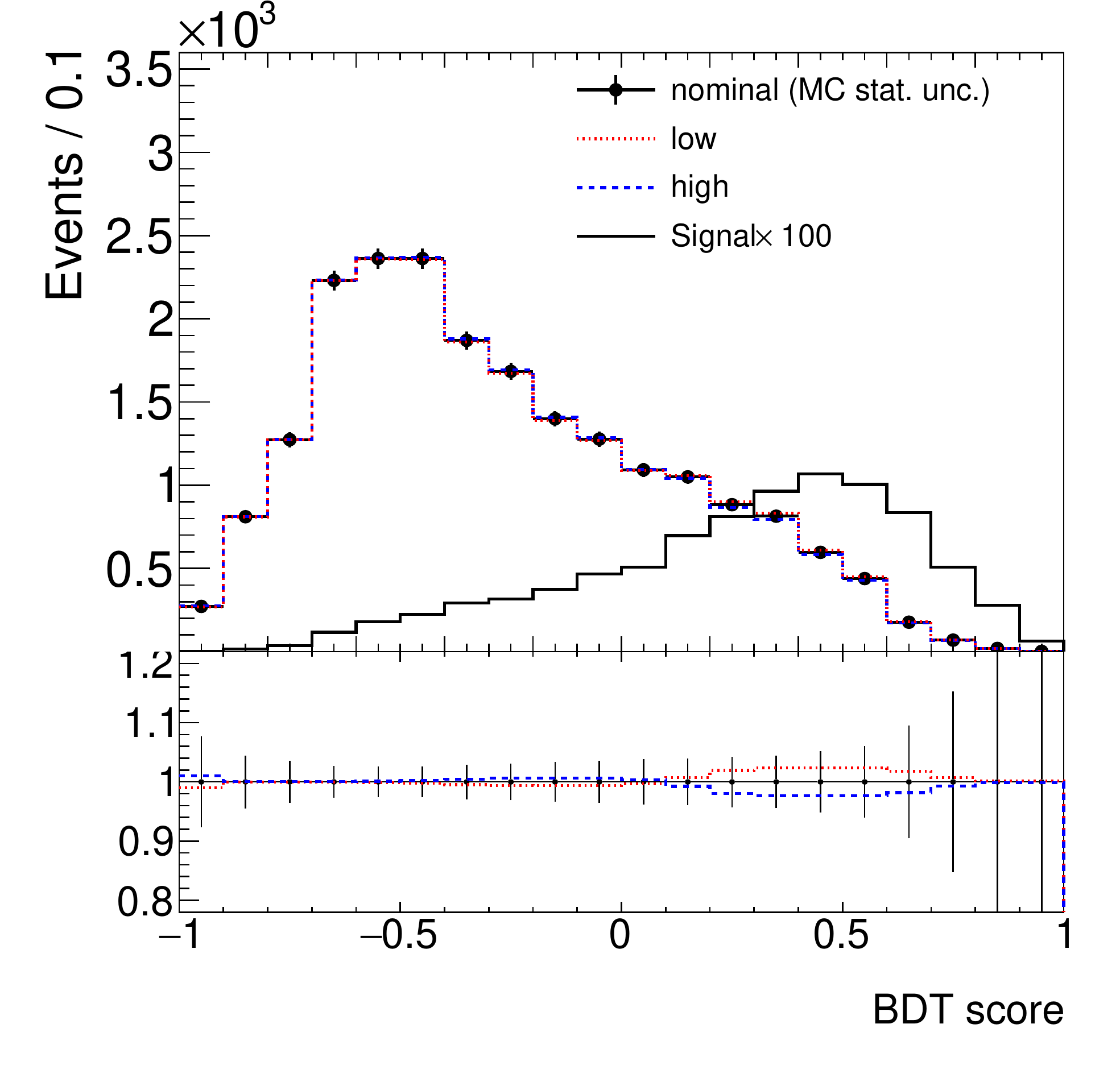}
	\includegraphics[width=0.35\textwidth]{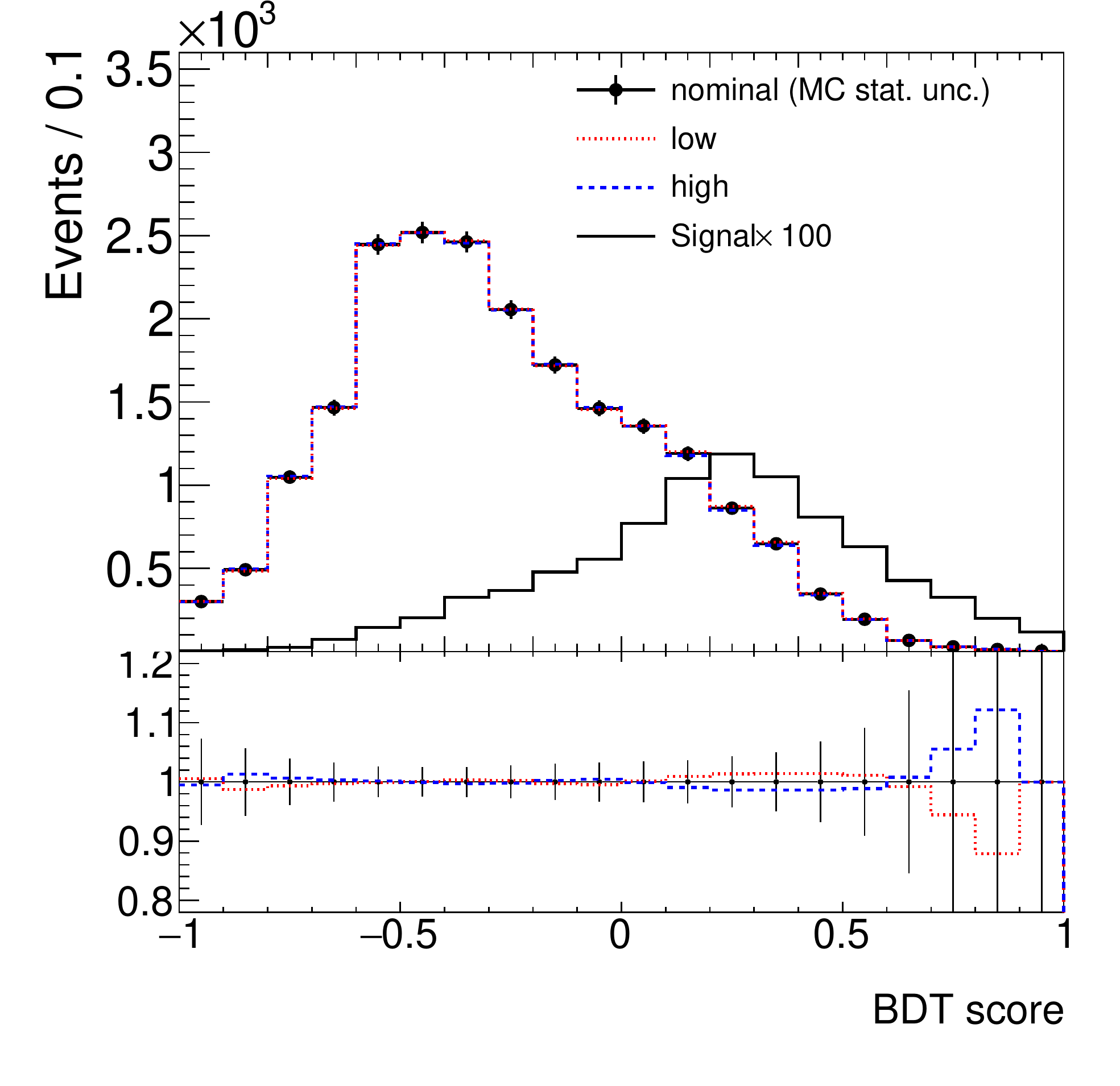}\\
	\includegraphics[width=0.35\textwidth]{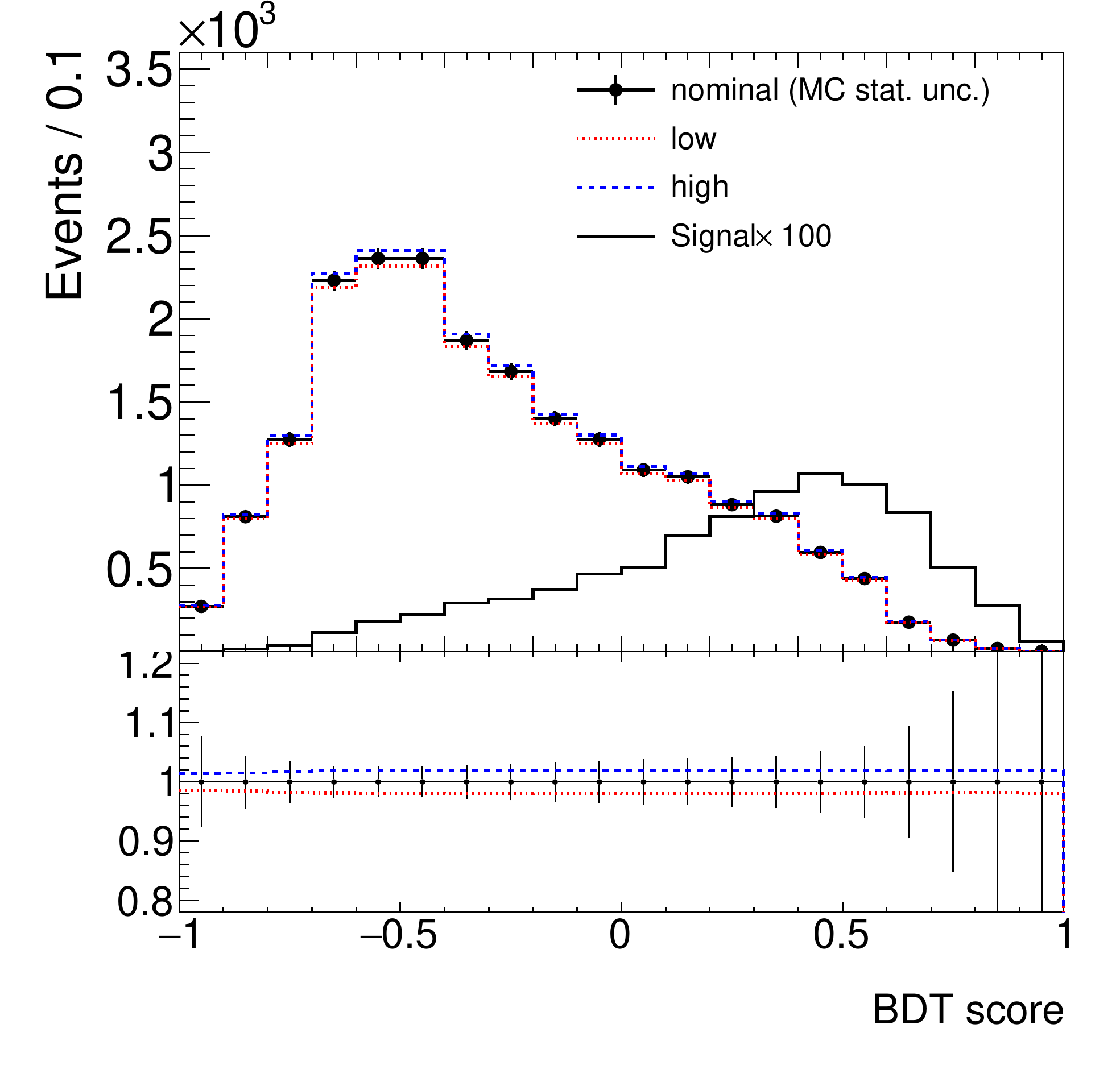}
	\includegraphics[width=0.35\textwidth]{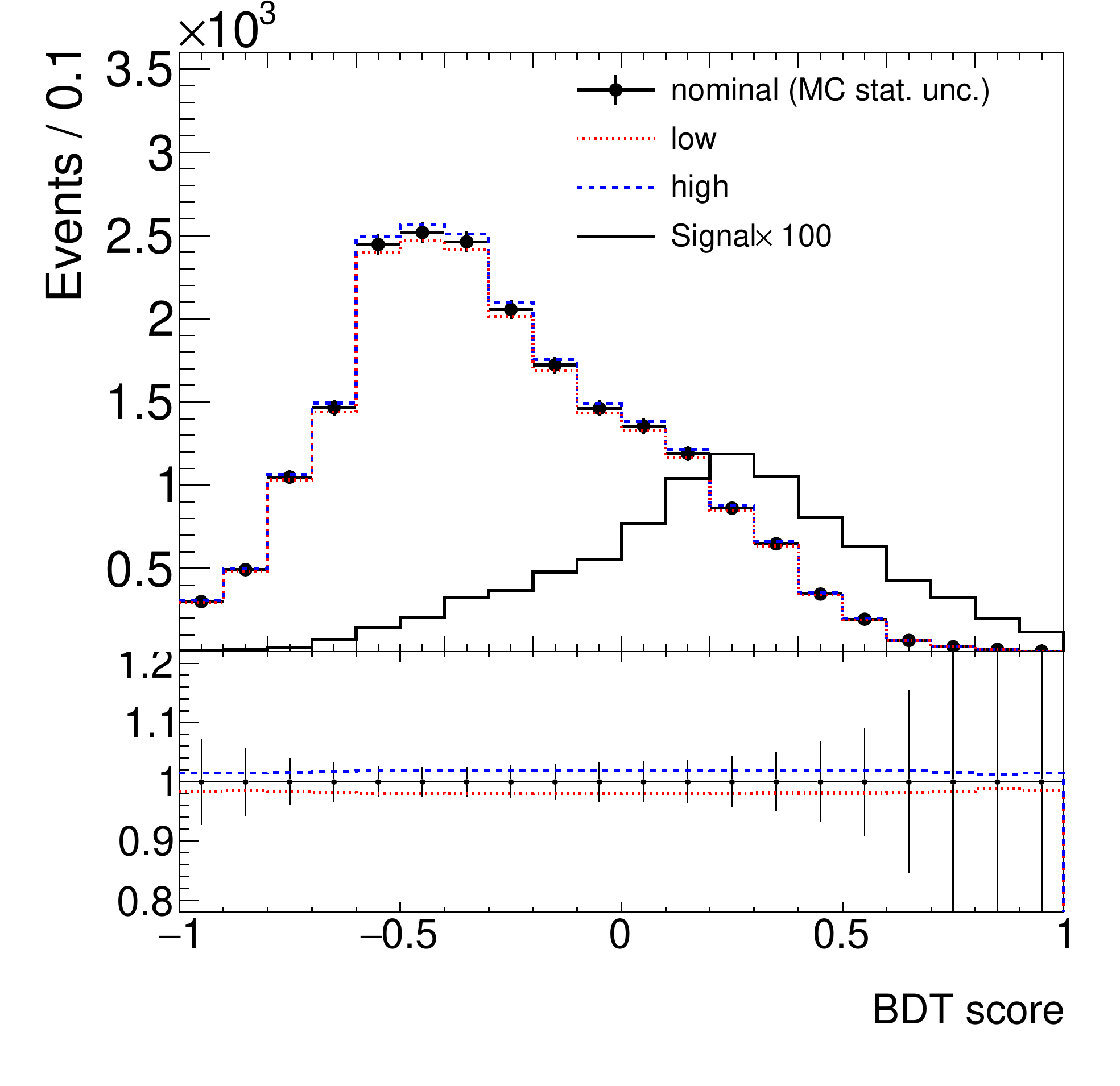}
	\caption{\label{fig:ex7_envelope_part2}
        (color online) Envelope plot for the systematical uncertainties. Left column: GradBDT, right column: QBDT7. From top to bottom, the systematical uncertainty source is lepton $\pT$ calibration, lepton ID efficiency, photon $\pT$ calibration and photon ID efficiency.
	}
\end{figure}

   The envelope plots are compared between GradBDT and QBDT7 in Fig.~\ref{fig:ex7_envelope_part1} and ~\ref{fig:ex7_envelope_part2}. The fit results are summarized in Table~\ref{tab:ex7_sig}. The same as in Case~I and II, the correlation between signal and the other nuisance parameters is generally reduced and QBDT7 gives the best significance. 
The signal strength uncertainty $\Delta\mu$ due to the systematical uncertainties in QBDT is 50\% of that in GradBDT.   
This case is to show that QBDT can take effect with the presence of many systematical uncertainty sources and seems to work better with more systematical uncertainties. 

Comparing three successive cases, we can see the significance for GradBDT and QBDT0 decreases with more systematical uncertainties included. This situation, however, is quite different for QBDTX which considers systematical uncertainties in training. On the contrary, the significance for QBDT7 with seven systematical items is actually higher than that for QBDT1 with only one systematical items. This is also reflected by the $\Delta\mu$ values in Table~\ref{tab:ex3_dmu}. 
The reason is that the correlation between the signal strength (POI) and the systematical uncertainties (nuisance parameters) or the correlation between different systematical uncertainties are different between GradBDT and QBDT. In Appendix~\ref{app:correlation}, we show how the correlation could be different for different ML algorithms and how it could make a difference.

\begin{table}
   \caption{\label{tab:ex7_sig} 
       Expected significance (unit: $\sigma$) expressed in number of standard deviation and post-fit uncertainty of the nuisance parameters. $\sigma_\theta$ is the post-fit uncertainty of the nuisance parameter $\theta$.
   }
   \begin{ruledtabular}
     \begin{tabular}{l | c c c}
         Significance ($\sigma$) & GradBDT & QBDT0 & QBDT7  \\
        \hline
	  Stat.-only fit & 0.89 & 0.90 & 0.91\\
	  Full fit & 0.66 & 0.68 & 0.83 \\
	  \hline
	  $\sigma_\theta(\tau \pT)$ & 0.57 & 0.59 & 0.59\\
	   $\sigma_\theta(\tau\ID)$ & 0.48 & 0.47 & 0.50 \\
            $\sigma_\theta(\MET)$ & 0.89 & 0.88 & 0.78 \\
	      $\sigma_\theta(l\pT)$ & 0.91 & 0.95 & 0.91\\
	      $\sigma_\theta(l\ID)$ & 0.99 & 0.99 & 0.99 \\
	$\sigma_\theta(\gamma \pT)$ & 0.85 & 0.95 & 0.86\\
	 $\sigma_\theta(\gamma\ID)$ & 0.94 & 0.93 & 0.94 \\
     \end{tabular}
   \end{ruledtabular}
\end{table}


\begin{figure}
   \includegraphics[width=0.32\textwidth]{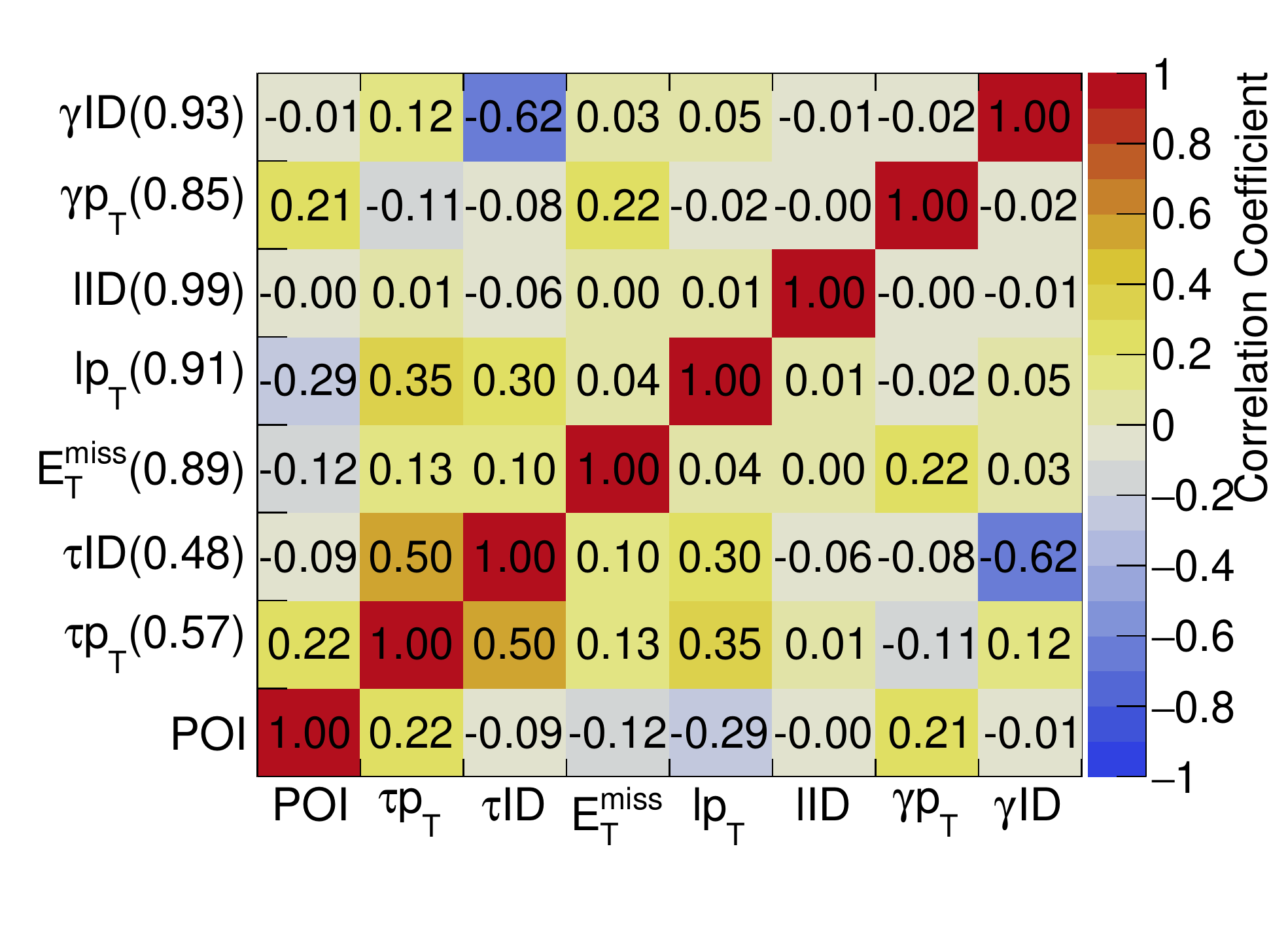}
   \includegraphics[width=0.32\textwidth]{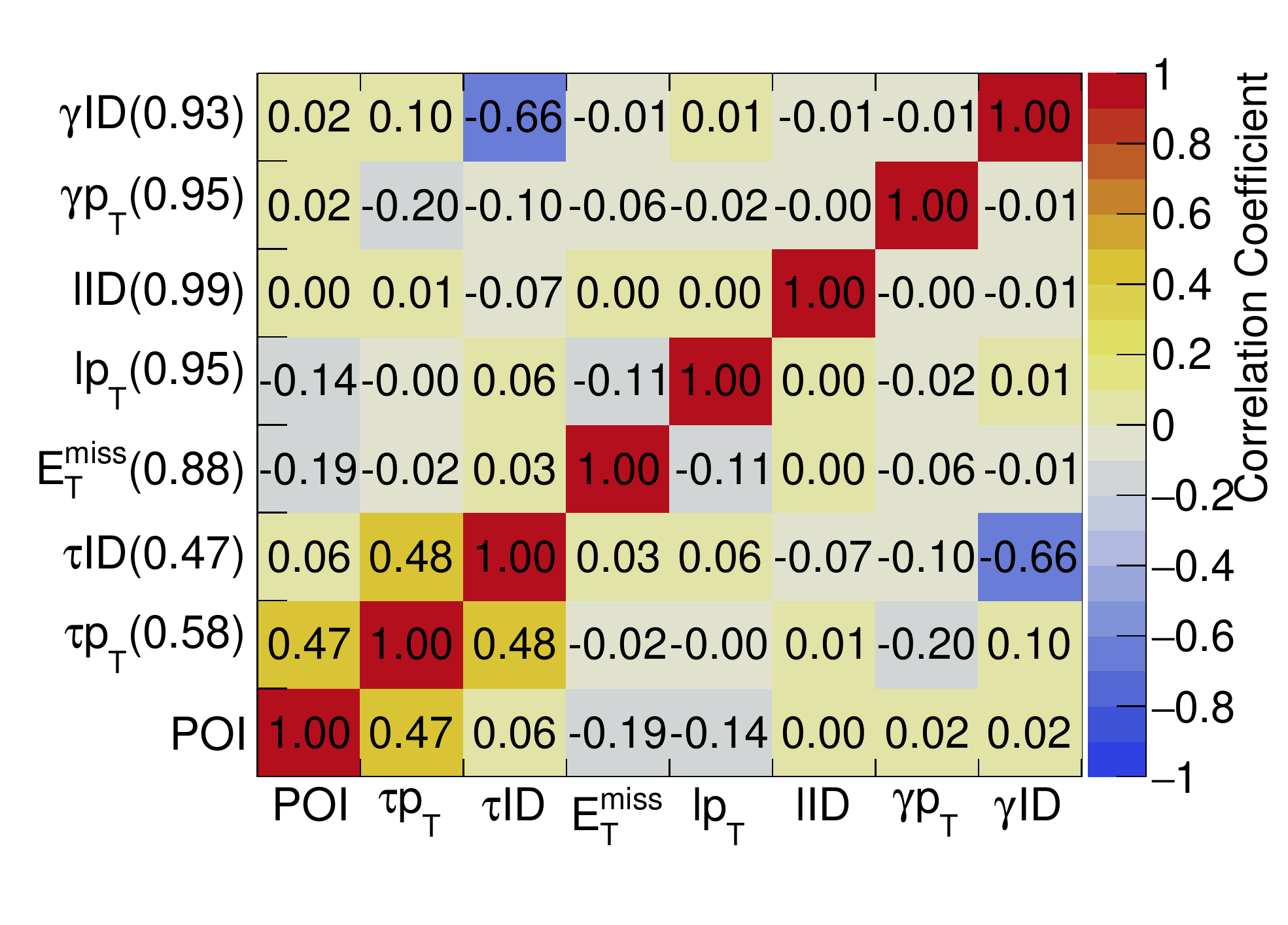}
   \includegraphics[width=0.32\textwidth]{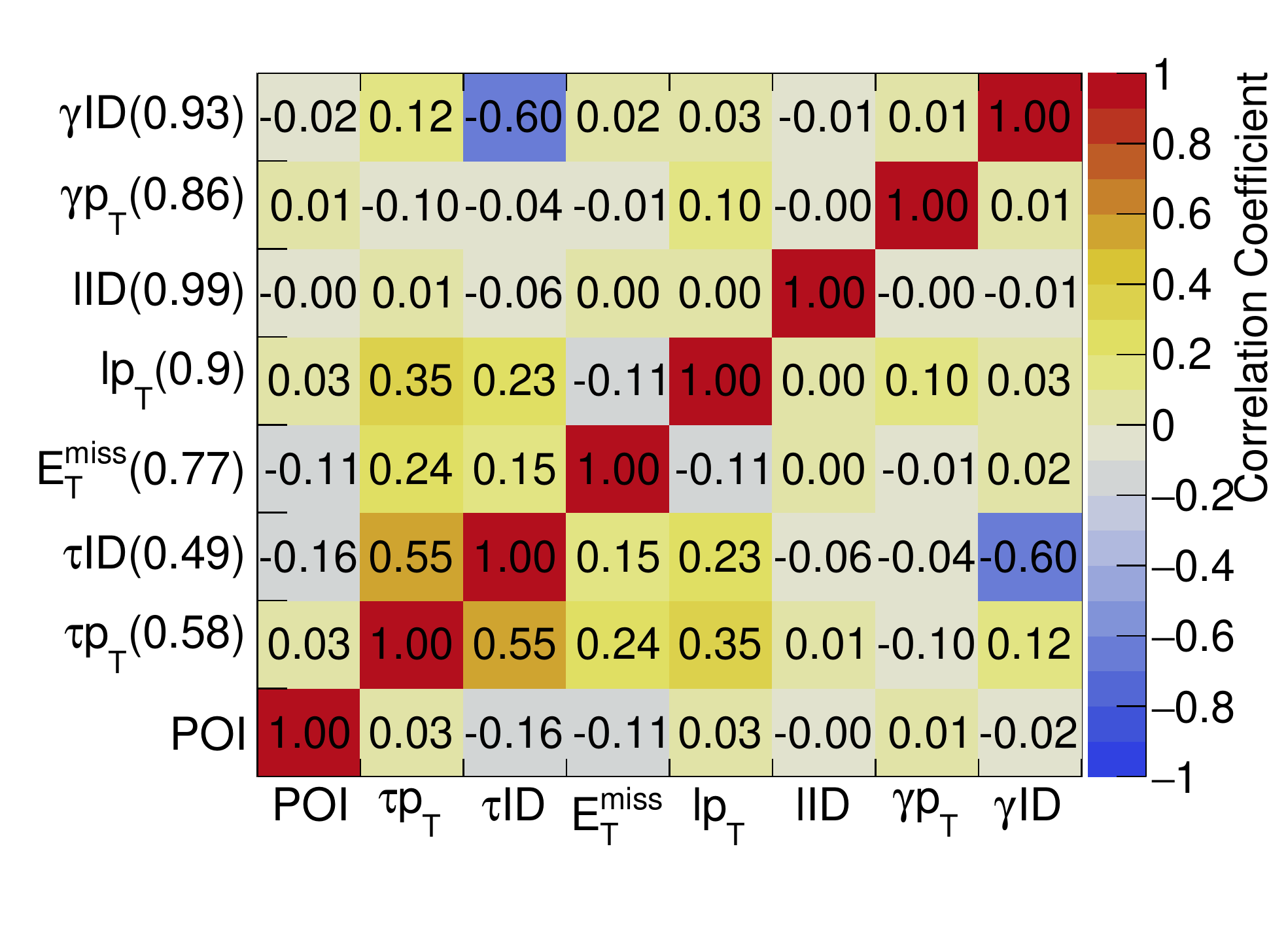}
   \caption{\label{fig:ex7_corr}
	Correlation matrix obtained from the fits. Left: GradBDT, middle: QBDT0 and right: QBDT7. 
   }
\end{figure}

\section{Summary}~\label{sec:summary}

In summary, a new boosting decision tree method, QBDT, is proposed to reduce the effect of the systematical uncertainties for the classification problem in high energy physics. The concept of significance in HEP plays two important roles in this method. One is that the node split in building a tree is determined by maximizing the significance. This is similar to the split criteria in GradBDT, where the node split is determined by minimizing the loss function. The other is to act
as the tree weight, which is part of the BDT score, and used to update the event weight for subsequent tree. This point is similar to the tree update rule in AdaBDT, where heavier weight is applied to the misclassified events. As the various systematical uncertainties can be conveniently included in the calculation of the total significance, QBDT is able to perform training with the systematical uncertainties. Taking a typical HEP search for the rare radiative Higgs decay $pp\to h+X
\to \gamma\tau^+\tau^-+X$ as example, it turns out that the correlation between the signal strength and the systematical uncertainty sources is reduced in QBDT. Therefore, QBDT gives a better signal significance. Based on the example, the contribution to the signal strength uncertainty from the systematical uncertainty sources in the QBDT method is 50--85~\% of that in the GradBDT method.
In this work, we only consider the usage of QBDT in classification. It might be applicable to regression problems as well. We will consider this possibility in the future.

\section{Acknowledgement}
I would like to thank Sinead Farrington for bringing this topic to my attention and thank Fang Dai for encouraging words. 

\begin{appendix}
    \section{Various significance representations with one systematical uncertainty}\label{app:significance}
In Case I presented in Sec.~\ref{sec:ex1}, we find that the nuisance parameter $\theta(\tau \pT)$ is over-constrained compared to the prior constraint with a similar degree for all BDT methods. Here we try to investigate the effect of the prior constraint. We consider a likelihood fit performed to a binned distribution of an observable with one systematical uncertainty source. Let $n_i$, $s_i$ and $b_i$ be the number of events in the $i$-th bin from data, signal predication and background estimation, respectively. The uncertainty of the background number of events due to the systematical source is denoted by $\Delta_i$ in the $i$-th bin.  With $\Nbins$ denoting the number of bins, the likelihood function is
\begin{equation}
\mathcal{L}(\mu,\theta) = \Pi_{i=1}^{\Nbins} P(n_i|\lambda_i(\mu,\theta)) \times G(\theta|0,A^2) \: .
\end{equation}
Here $P(n|\lambda) \equiv \frac{\lambda^n}{n!}e^{-\lambda}$ is the Poisson distribution with the expectation value $\lambda$, and $G(\theta|0,A)\equiv\frac{1}{\sqrt{2\pi}}e^{-\frac{\theta^2}{2A^2}}$ is the Gaussian distribution to introduce the constraint on the systematical uncertainty with the nuisance parameter $\theta$ and a fixed parameter $A$ to control the degree of constraining the systematical uncertainty size. In HEP, systematical uncertainty is usually estimated in an independent measurement and introduced to other analyses with $A$ renormalized to be 1.  If the post-fit uncertainty of $\theta$, $\sigma(\theta)$, is less than 1,  we say the systematical uncertainty is over-constrained and may worry about it.
However, if the observable, sensitive to the signal, is also sensitive to the systematical uncertainty or if the independent estimation of the systematical uncertainty is very conservative, it is expected that over-constraining happens. Therefore we write down $A$ explicitly for further discussions. Letting $\mu$ denote the parameter of interest to control the signal strength, the expected number of events in the $i$-th bin is $\lambda_i(\mu,\theta) \equiv \mu s_i + b_i + \theta \Delta_i$.  Ignoring the irrelevant constant terms, the log likelihood function is
\begin{equation}
\log\mL(\mu, \theta) = - \frac{\theta^2}{2A^2}+\sum_{i=1}^{\Nbins} n_i\log\lambda_i(\mu,\theta) - \lambda_i(\mu,\theta) \:.
\end{equation}

The best estimation of  $\mu$ and $\theta$ is obtained by maximizing the likelihood function, namely, $\frac{\partial \log\mL}{\partial \mu}= 0$ and $\frac{\partial \log\mL}{\partial \theta}= 0$. Using $\hatmu$ and $\hatthe$ to denote the best-fit values, they satisfy 
\begin{eqnarray}
&& \sum_{i=1}^{\Nbins}\frac{n_i}{\lambda_i(\hatmu,\hatthe)}s_i-s_i = 0 \:, \label{eq:mu}\\
&& -\frac{\hatthe}{A^2}+\sum_{i=1}^{\Nbins}\frac{n_i}{\lambda_i(\hatmu,\hatthe)}\Delta_i-\Delta_i=0  \: . \label{eq:theta}
\end{eqnarray}
To get the expected significance for the signal strength $\mu=1$, we consider an Asimov data set with $n_i = 1\times s_i+b_i$ for all $i$'s and calculate the likelihood under the null (background only) and alternative (signal plus background) hypothesis. For the latter hypothesis, the best-fit values are simply $\hatmu=1$ and $\hatthe=0$. For the null hypothesis, we need to solve Eq.~\ref{eq:theta} in which we should fix $\hatmu=0$.  Assuming $|\Delta_i|<<b_i$, we have
\begin{equation}
\frac{1}{\lambda_i(0,\theta)} = \frac{1}{b_i(1+\frac{\theta\Delta^i}{b_i})} \approx \frac{1}{b_i}(1-\frac{\theta\Delta_i}{b_i}) \: .
\end{equation}
Using the approximation above, the best-fit $\theta$ (denoted by $\hat{\hatthe}$) is found to be (to the order of $|\Delta_i|/b_i$)
\begin{equation}
\hat{\hatthe}(A) \approx \frac{\sum_{i=1}^{\Nbins}\frac{n_i}{b_i}\Delta_i -\Delta_i}{\frac{1}{A^2}+\sum_{i=1}^{\Nbins}\frac{n_i\Delta_i^2}{b_i^2} } 
=  \frac{\sum_{i=1}^{\Nbins}\frac{s_i}{b_i}\Delta_i }{\frac{1}{A^2}+\sum_{i=1}^{\Nbins}(1+\frac{s_i}{b_i})\frac{\Delta_i^2}{b_i} } \: .
\end{equation}
With $\hatmu$, $\hatthe$ and $\hhatthe$, we can calculate the likelihood ratio, $Q$,
\begin{equation}
Q(A) \equiv 2\ln\frac{\mL(\hatmu,\hatthe)}{\mL(0,\hat{\hatthe}(A))} = \sum_{i=1}^{\Nbins}2\left[(s_i+b_i)\log\frac{s_i+b_i}{b_i+\hhatthe(A)\Delta_i}-(s_i-\hhatthe\Delta_i)\right] + \frac{\hhatthe^2(A)}{A^2}
\end{equation}
and use $Z_Q(A) \equiv \sqrt{Q(A)}$ to measure the total significance. $A=1$ corresponds to including the prior constraint into the likelihood fit while $A=+\infty$ means not considering the prior constraint.

For comparison, we introduce two more quantities, $Z_0$ and $Z_{\chi^2}$, which can also measure the significance with systematical uncertainty. 
\begin{eqnarray}
&& Z_0 = \sqrt{\sum_{i=1}^{\Nbins}2\left[(s_i+b_i)\ln\frac{(s_i+b_i)(b_i+\Delta_i^2)}{b_i^2+(s_i+b_i)\Delta_i^2}-\frac{b_i^2}{\Delta_i^2}\ln(1+\frac{s_i\Delta_i^2}{b_i(b_i+\Delta_i^2)})\right]} \: , \label{eq:ZQ}\\
&& Z_{\chi^2} = \sqrt{\sum_{i=1}^{\Nbins}\frac{s_i^2}{b_i+\Delta_i^2}} \: .\label{eq:Zchi2}
\end{eqnarray}
Obviously, $Z_0$ is an extension to the case of multiple bins from Eq.~\ref{eq:Q}. It is derived assuming the uncertainties in all bins are independent, which assumption does not hold in reality. $Z_0$ is close to $Z_{\chi^2}$ if $s_i<<b_i$ and $|\Delta_i|<<b_i$.  

Table~\ref{tab:ex123_Zs} summarizes various representations of significance with a single systematical uncetainty for all BDT score distributions. In Case~I presented in Sec.~\ref{sec:ex1}, $\theta(\tau\pT)$ is found to be over-constrained, but we find that $Z_Q(1)$ and $Z_Q(+\infty)$ are very close. This means the constraining comes from the observable itself. It confirms that the BDT score is actually sensitive to $\tau \pT$ calibration because of the input variables.
For completeness, we also calculate the significance for other systematical uncertainty sources. If multiple systematical uncertainties are present, there is no simple analytical expression to represent a significance reasonably. One could use the figure of merit~\cite{victor,inferno}, $\Delta\mu/\mu$ (relative uncetainty of the signal strength), where $\Delta\mu$ considers all systematical uncertainties. This is just what Table~\ref{tab:ex3_dmu} shows as we have seen in the main text.

\begin{table}
   \caption{\label{tab:ex123_Zs} 
  Various representations of significance}
   \begin{ruledtabular}
     \begin{tabular}{l | l | l  l l l}
	Syst. Unc. & BDT & $Z_{\chi^2}$ & $Z_0$ & $Z_Q(1)$ & $Z_Q(+\infty)$ \\
	  \hline
	   &GradBDT & 1.20 & 1.17 & 1.16 & 1.15 \\
	   $\tau \pT$ & QBDT0 & 1.20 & 1.16 & 1.17 & 1.15 \\
	    &QBDT1 & 1.09 & 1.07 & 1.26 & 1.26 \\
	     &QBDT3 & 1.29 & 1.26 & 1.24 & 1.23 \\
	   &QBDT7 & 1.24 & 1.21 & 1.39 & 1.39 \\ 
	   \hline
	   	   &GradBDT & 1.23 & 1.20 & 1.31 & 1.31 \\
	   $\tau$ID & QBDT0 & 1.29 & 1.25 & 1.35 & 1.34 \\
	     &QBDT3 & 1.26 & 1.23 & 1.33 & 1.33 \\
	   &QBDT7 & 1.31 & 1.28 & 1.36 & 1.36 \\ 
	   \hline
	   	   &GradBDT & 1.42 & 1.39 & 1.33 & 1.31 \\
	   $\MET$ & QBDT0 & 1.50 & 1.46 & 1.42 & 1.36 \\
	   &QBDT3 & 1.41 & 1.38 & 1.40 & 1.37 \\
	   &QBDT7 & 1.45 & 1.42 & 1.48 & 1.47 \\ 
	   \hline
	   	   &GradBDT & 1.40 & 1.38 & 1.27 & 1.16 \\
	  $l\pT$ & QBDT0 & 1.49 & 1.46 & 1.44 & 1.36 \\
	   &QBDT7 & 1.48 & 1.45 & 1.49 & 1.49 \\ 
	   \hline
	   	   &GradBDT & 1.46 & 1.44 & 1.43 & 1.31 \\
	   $l$ID & QBDT0 &  1.51 & 1.47 & 1.46 & 1.34 \\
	   &QBDT7 & 1.52 & 1.49 & 1.48 & 1.36 \\ 
	   \hline
	   	   &GradBDT & 1.40 & 1.38 & 1.31 & 1.23 \\
	   $\gamma\pT$ & QBDT0 & 1.50 & 1.46 & 1.44 & 1.36 \\
	   &QBDT7 & 1.47 & 1.44 & 1.49 & 1.48 \\ 
	   \hline
	   	   &GradBDT & 1.40 & 1.38 & 1.32 & 1.31 \\
	   $\gamma$ID & QBDT0 & 1.44 & 1.41 & 1.36 & 1.34 \\
	   &QBDT7 & 1.46 & 1.43 & 1.38 & 1.36 \\ 
     \end{tabular}
   \end{ruledtabular}
\end{table}

\section{Correlation evolution in QBDTX}\label{app:correlation}
In this section, let us investigate the correlation evolution in QBDTX. Firstly let us show that the correlation can evolve in different ML algorithms. Then we present an example to show how the correlation will make a difference.

For any typical ML algorithm, it is basically a function mapping the input variables $\vx=(x_1,x_2,\cdots)$ to a real number $y$, BDT score for BDT methods. We denote it by $y=f(\vx)$. For a class of events like signal or background denoted by $A=S$ or $B$, $x_1,x_2,\cdots$ may have a complicated joint PDF, $q_A(\vx)$. This will lead to a PDF of BDT score, $p_A(y)$, via the mapping $f$. 
For different algorithms, $q_A(\vx)$ is the same, while $y=f(\vx)$ and $q_A(y)$ are different. The various systematical uncertainties will affect $q_A(\vx)$ and thus affect $p_A(y)$. So $p_A(y)$ is extended to be $p_A(y,\vtheta)$ with $\vtheta$ denoting a set of nuisance parameters, $\theta_j (j=1,2,\cdots,\Nsysts)$. The unbinned likelihood function can be reconstructed as~\cite{histofactory}
\begin{equation}
    \mL(\mu,\vtheta) = P(n_D|\mu n_S+n_B)\times \Pi_{i=1}^{\Nevents}\frac{\mu n_Sp_S(y_i,\vtheta) + n_B p_B(y_i, \vtheta)}{\mu n_S + n_B} \times \Pi_{j=1}^{\Nsysts}G(\theta_j|0,1)\:,
\end{equation}
where $n_{D/S/B}$ represents the total number of data events, signal events and background events, respectively; $P(n|\lambda)$ and $G(\theta|0,1)$ are Poisson and normal distribution functions as defined before. 

The correlation coefficient between signal strength $\mu$ and systematical uncertainty $\theta_i$ or between different systematical uncertainties $\theta_i$ and $\theta_j$ can be obtained by evaluating the Hessian matrix at the best-fit values. It involves the second-order derivatives of the logarithmic likelihood function and of-course very complicated. An analytic approximate expression is available in Ref.~\cite{constraint_xia}, which result has been used in QBDT. 
Without going into details, the correlation can be different for different ML algorithms via  $p_A(y,\vtheta)$ in the likelihood function, though we cannot show the exact evolution mechanism behind either QBDT or GradBDT.

Then let us take the uncertainty of the total number of background events, $\sigma_b$, as example to show how the correlation works. We can calculate $\sigma_b$ according to Eq.~\ref{eq:sigmab0}. If not considering the correlations, we have (from Eq.~\ref{eq:sigmab0})
\begin{equation}
    \sigma_b^2 = \sum_{i=1}^{\Nsysts}\left(\frac{\partial b}{\partial\theta_i}\sigma_{\theta_i}\right)^2 \:,
\end{equation}
which increases with the number of systematical uncertainties. However the correlation contribution (second term in Eq.~\ref{eq:sigmab0}) could be negative.
We actually have all ingredients for the calculation, namely, the post-fit uncertainty shown in Table~\ref{tab:ex1_sig},~\ref{tab:ex3_sig} and ~\ref{tab:ex7_sig}, and the correlation matrices shown in Fig.~\ref{fig:ex1_corr}, ~\ref{fig:ex3_corr} and ~\ref{fig:ex7_corr}. 

The calculation is straightforward and the numerical results are summarized in Table~\ref{tab:ex123_sigmab}. The results without considering the correlations are shown by the numbers in the brackets (there is no correlation terms for Case~I as there is only one systematical uncertainty). Comparing three cases and three BDT setups, $\sigma_b$ increases with more systematical uncertainties in GradBDT and QBDT0. But for QBDTX, $\sigma_b$ in Case~III is even smaller than that in
Case~II. This example is to show that the correlation does make a difference. Readers who are interested in the correlation contribution to the signal strength uncertainty can refer to Ref.~\cite{constraint_xia}. It seems that the reduction of the correlation between signal strength and systematical uncertainties is the only solution to reduce the impact on POI uncertainty (thus to increase the significance). Therefore, it is probably true that we will see the same correlation reduction in any effective ML algorithm with systematical uncertainties in the future. 

\begin{table}
   \caption{\label{tab:ex123_sigmab} 
  Uncertainty on the total number of background events. The results without considering the correlations are shown by the numbers in the brackets.}
   \begin{ruledtabular}
     \begin{tabular}{l | l l l }
	   & GradBDT & QBDT0 & QBDTX\\
         \hline
         Case~I & 220.11 & 231.20 & 218.77 \\
         Case~II & 257.16(470.94) & 258.94(490.37) & 254.70(528.96) \\
         Case~III & 259.92(704.53) & 260.22 (698.41) & 247.85(716.10) \\
     \end{tabular}
   \end{ruledtabular}
\end{table}

\end{appendix}

\end{document}